\documentclass{emulateapj}
\usepackage{times}
\usepackage{amsmath}
\usepackage{graphicx}

\slugcomment{To appear in ApJ}

\shorttitle{H$\alpha$ Pulsar Bow Shocks} 
\shortauthors{Brownsberger \& Romani}

\begin{document}

\title{A Survey for H$\alpha$ Pulsar Bow Shocks} 

\author{Sasha Brownsberger \& Roger W. Romani\altaffilmark{1}}
\affil{Dept. of Physics/KIPAC, Stanford University, Stanford, CA 94305-4060}
\altaffiltext{1}{Visiting Astronomer, Kitt Peak National and Cerro Tololo Inter-American 
Observatories, which are operated by the Association of Universities for Research in Astronomy 
(AURA) under cooperative agreement with the National Science Foundation. 
The WIYN observatory is a joint facility of the University of Wisconsin-Madison, 
Indiana University, Yale University, and the National Optical Astronomy Observatory.
The SOAR telescope is a joint project of: Conselho Nacional de Pesquisas Cientificas e 
Tecnologicas CNPq-Brazil, The University of North Carolina at Chapel Hill, Michigan State 
University, and the National Optical Astronomy Observatory.}
\email{rwr@astro.stanford.edu, sashab@stanford.edu }

\begin{abstract}

	We report on a survey for H$\alpha$ bow shock emission around 
nearby $\gamma$-detected energetic pulsars. This survey adds three
Balmer-dominated neutron star bow shocks to the six previously confirmed examples.
In addition to the shock around {\it Fermi} pulsar PSR J1741$-$2054,
we now report H$\alpha$ structures around two additional $\gamma$-ray pulsars,
PSR J2030+4415 and PSR J1509$-$5850. These are the first known examples of H$\alpha$ nebulae
with pre-ionization halos. With new measurements, we show that a simple analytic model
can account for the angular size and flux of the bow shocks' apices. The latter, in particular,
provides a new pulsar probe and indicates large moments of inertia and smaller distances
than previously assumed, in several cases. In particular we show that the re-measured PSR J0437$-$4715 shock
flux implies $I = (1.7\pm 0.2) \times 10^{45}/(f_{HI} {\rm sin}i) {\rm g\,cm^2}$. We also derive
a distance $d\approx 0.72$\,kpc for the $\gamma$-ray only pulsar PSR J2030+4415 and revised 
distances for PSRs J1959+2048 (1.4\,kpc) and J2555+6535 ($\sim 1$\,kpc), smaller than the 
conventional DM-estimated values.
Finally we report upper limits for 94 additional LAT pulsars. An estimate of the
survey sensitivity indicates that for a warm neutral medium filling factor 
$\phi_{WNM}\sim 0.3$ there should be a total of $\sim$ nine H$\alpha$ bow shocks in
our LAT-targeted survey; given that seven such objects are now known, a much 
larger $\phi_{WNM}$ seems problematic.
\end{abstract}


\section{Introduction}

	Pulsar wind nebulae (PWNe) have been observed {\it via} X-ray synchrotron
radiation for many sources over many years \citep[][and references therein]{kp10}. These data
probe the bulk energetics of the pulsar wind and, with high
quality images from the {\it Chandra} X-ray observatory, reveal
termination shock structure in the form of tori and jets. When the
pulsar is subsonic, as in their parent supernova remnant, these X-ray structures
can show high symmetry about the spin axis. However, when the pulsar escapes to the cooler
external interstellar medium (ISM), the pulsar motion is generally supersonic and the termination
shocks are ram pressure confined due to the pulsar motion. This asymmetry can
distort the relativistic PWN termination shock. Thus it was
particularly interesting when an H$\alpha$ bow shock was discovered around the
black widow millisecond pulsar PSR B1957+20 \citep{kh88}, both because this
structure showed high symmetry about the pulsar velocity axis and because 
the Balmer line emission allows kinetic study of the shock structure.
These bow shocks
are non-radiative, with the pulsar velocity ensuring that the shocked
ISM flows well down-stream before it cools. Non-radiative shocks show extreme
Balmer domination and the velocity structure of these lines turns out to
be a sensitive probe of the shock physics, including information on the
up-stream ISM, the post-shock flow and even supra-thermal particle acceleration
\citep{waget09}. Pulsar bow shocks have an interesting range of speeds (low compared to most
supernova remnants) and shock obliquities, so observation of these nebulae
provide an excellent opportunity to study non-radiative shock physics.

	Given this interest, a number of searches have been mounted for similar pulsar nebulae
\citep[e.g.][]{crl93,ganet02}, with some limited success. Before the present
search there were six confirmed H$\alpha$ pulsar bow shocks reported in the 
literature. A shock around PSR J1741$-$2054 was discovered in the initial phase 
of the present search \citep{ret10}.  A few other candidates have been presented in 
the literature, but clearly these objects are quite rare.

	We report here on a new sensitive survey of a large number of nearby 
energetic pulsars. We have tested this sensitivity with images
of known bow shocks. Despite the modest exposures, these are in general
comparable with or superior to earlier images in the literature, and show additional
interesting H$\alpha$ structure. Our survey has added three new bow shock detections
(two reported here for the first time).  We use the new measurements, and a re-assessment
of the other known bow shocks, to illustrate the dependence on pulsar properties.
In addition we derive quantitative upper limits
on the H$\alpha$ fluxes from the non-detections. This is now a large enough 
sample that it can be used to evaluate, heuristically, the detection probability's
dependence on pulsar parameters. Also, for a basic assumed model,
the limits provide an interesting spot sample of the neutral ISM within a few
kpc of the Sun.

\section{Bow Shock Basics}

	One can estimate the spin-down luminosity of a rotation-powered pulsar
as ${\dot E} = 4\pi^3 I {\dot P}/P^3$. Conventionally one assumes $I=10^{45} {\rm g\, cm^2}$.
However, \citet{vKK11} have estimated the mass of PSR J1959+2048
as $2.4\pm0.16M_\odot$. To accommodate such a large mass, the equation of state must be
stiff, and the moment of inertia correspondingly large. \citet{ls05} find that for
stiff equations of state the moment of inertia is approximately
$$
I \approx (0.8 - 1.2) \times 10^{45} M^{1.5} {\rm g\,cm^2}
\eqno (1) 
$$
where the mass is measured in solar masses and the prefactor covers the range allowed
by the equations of state that can accommodate PSR J1959+2048's mass (from AP4
to the most extreme MS0, in their compilation). To be conservative, we adopt here the
smaller coefficient in this estimate and for sources without published neutron star
mass measurements we will assume $1.4M_\odot$ ($I_{45}=1.3$) for young
pulsars and $1.8M_\odot$ ($I_{45}=1.9$) for strongly recycled millisecond pulsars.
Our analysis is not very sensitive to this parameter. A more detailed sum
could include a mass distribution \citep[e.g.][]{ozet12}, but this should be
computed for the short period, very energetic LAT sample, including black-widow
type pulsars.

	For radio pulsars we have a distance estimate $d$ from the dispersion delay or, 
less often, but much more reliably, a
parallax measurement from pulse timing or interferometry. These are often supplemented
by proper motion estimates $\mu$, so that we can infer a transverse space velocity
$v = 4.75 \mu_{\rm masy} d_{kpc}/{\rm sin}i\, {\rm km\,s^{-1}}$, where $\mu_{\rm masy}$
is measured in milli-arcseconds/year, $d_{kpc}$ is the distance in kiloparsecs
and $i$ the angle between ${\vec v}$ and the Earth line-of-sight. This
velocity can be used to correct the observed spindown for the
\citet{shk70} effect 
$$
{\dot P}={\dot P}_{obs} - 2.43\times 10^{-21} P\, \mu_{\rm masy}^2 d_{kpc} 
\eqno(2)
$$
which can be substantial for high velocity millisecond pulsars. We ignore here the smaller
perturbations to ${\dot P}_{obs}$ caused by acceleration in the Galactic potential. Equations
(1) and (2) improve our estimate of the pulsar spin-down luminosity ${\dot E}$.

	An H$\alpha$ bow shock forms when a pulsar of luminosity 
$10^{34} {\dot E}_{34} {\rm erg\, s^{-1}}$ travels supersonically 
through a partly neutral medium of density $\rho = \gamma_H m_p n_{\rm H}$.
Along ${\vec v}$ the pulsar wind and ISM ram pressure balance at the contact discontinuity
at standoff angle
$$
\theta_0= ({\dot E}/4\pi c \rho v^2 )^{1/2}/d .
\eqno(3)
$$
The H$\alpha$-producing forward shock lies upstream, at an apex distance of 
$\theta_a\approx (1.3-1.5)\times \theta_0$ \citep{arc02,buc02}, so we infer a characteristic angular
scale for the bow shock apex of
$$
\theta_a= 1.3 \theta_0 = 19.7^{\prime\prime} [{\dot E}_{34} {\rm sin}^2 i/
(n_{\rm H}\mu_{\rm masy}^2) ]^{1/2}\, d_{kpc}^{-2},
\eqno(4)
$$ 
using $\gamma_H=1.37$ to convert H density to total density.  In the
simple case of a spherically symmetric wind \citet{wil96} has provided a convenient
analytic description of the contact discontinuity stand-off for the thin-shock case
$$
r(\phi) = r_0 [3 (1-\phi {\rm cot} \phi)]^{1/2}/{\rm sin}\phi
\eqno(5)
$$
where $\phi$ is measured from the direction of motion with respect to the ambient ISM.
As noted by \citet{ret10}, for such a wind the projected angle to the bow shock limb
at the apex is nearly independent of $i$, as long as $i$ is not very small. Numerical
simulation \citep[eg.][]{buc02} shows that finite pressure causes the forward shock
to thicken slightly along the bow shock; in the vicinity of the pulsar this is
well approximated by increasing the transverse size to
$$
r_\perp=1.25 \times 1.3\,r(\phi) {\rm sin}(\phi)
\eqno(6)
$$
over a few$\times\, \theta_a$.

	Further downstream, bow shocks are visible to 100$\times\theta_a$ and a wide range 
of structures are seen, often with multiple cavities or bubbles, suggesting either
ISM density variations or ${\dot E}$ instabilities \citep{vKI08}. Also,
\citet{wil00} extended the thin shock analysis to give expressions for the more general case
of an axisymmetric wind, possibly misaligned with the pulsar motion. Indeed
a few bow shock limb shapes are best fit including an equatorial concentration
for the pulsar wind \citep{viget07,ret10}.

	However, near the apex all observed bow shocks have a similar shape, allowing reasonable
estimates for the standoff scale and flux by measuring in an aperture following equations (5)
and (6).
We define the apex zone from $\theta_a$ ahead of the pulsar to $-2\theta_a$ behind.
Study of this zone, when resolved, allows useful comparison between bow shocks 
(although as noted for a few cases detailed fits to the limb give significant constraints 
on velocity inclination $i$ and wind asymmetry). From Equation (5) the apex region
extends 2.77$r_a$ transverse to the direction of motion, and so, including our approximation
for the forward shock thickness, the apex sweeps up a flux of 
$$
{\dot N} = \pi (3.45\, r_a)^2 v \, n_H f_{HI}
\eqno(7)
$$
neutral H atoms per second, where $f_{HI}$ is the neutral fraction. These neutrals pass into the
heated, shocked ISM where they suffer collisional excitation and charge 
exchange before ionization at a distance 
$l \approx v/(R_I \, n) \sim (1-3) \times 10^{15} v_7n^{-1}{\rm cm}$, with the
ionization rate $R_I \sim (0.3-1) \times 10^{-8} {\rm cm^3\,s^{-1}}$ in the bow shock velocity
regime $v=10^7v_7{\rm cm\,s^{-1}}$. Note that 
$$
l/r_a \approx 10^{-4} (\mu_{\rm masy} d_{kpc}/{\rm sin}i)^2 (n_H\, {\dot E}_{34})^{-1/2}
\eqno(8)
$$
so $l \ll r_a$ except for the highest velocity low luminosity pulsars; we expect that the 
H$\alpha$-emitting layer is thin and that the ISM is generally fully ionized before
the contact discontinuity.
For non-radiative shocks in supernova, the rule of thumb is $\epsilon_{\rm H\alpha} \sim 0.2$ emitted
H$\alpha$ photons per incoming neutral \citep{ray01}. However our pulsar bow shocks have
relatively low velocities $v_7=1-10$, and the computations of \citet{hm07}
indicate increased H$\alpha$ yield for $v<10^3 {\rm km\,s^{-1}}$
$$
\epsilon_{{\rm H}\alpha} \approx 0.6 v_7^{-1/2}
\eqno(9)
$$
if the post-shock electrons and ions equilibrate via plasma-wave interactions. If out of
equilibrium, the yield appears to be lower at small velocity with $\epsilon_{{\rm H}\alpha} \approx
0.04 v_7^{3/4}$. 
Together these estimates give an H$\alpha$ flux at Earth of
\begin{eqnarray*}
f_{{\rm H}\alpha} &=& {\dot N} \epsilon_{{\rm H}\alpha}/(4\pi d^2) \qquad\qquad\qquad\qquad\qquad\qquad\qquad (10)\\
 &=& 3.5 \times 10^{-2} {\dot E}_{34} {\rm sin}^{3/2}i\,f_{\rm HI} \mu_{\rm masy}^{-3/2} d_{kpc}^{-7/2}
{\rm cm^{-2}\,s^{-1}} 
\end{eqnarray*}
where we assume post-shock $e^-$-ion equilibration and have not
explicitly included the dependence of ${\dot E}_{34}$ on velocity or neutron star mass.
One should note that the apex H$\alpha$ flux is essentially independent of the density $n$
as long as the ISM is fully ionized before the momentum balance at the contact
discontinuity.

	The LAT-selected pulsars are energetic and in nearly all cases have very small $l/r_a$
(equation 8).  However, the two non-LAT pulsars J1856$-$3754 and J2225+6535 are
high velocity, low ${\dot E}$ objects and we expect incomplete ionization at the bow shock
apex. This is case A of \citet{bb01}. Here we expect the momentum balance to be determined
by the post shock ionized fraction. This will be a combination of the upstream ionization
fraction $x_i=(1-f_{\rm HI})$ and incomplete post-shock ionization prior to the contact discontinuity 
with fraction $\approx 0.3\,r_a/l$. If
$$
x_i > 340 ({\dot E}_{34} n)^{1/3} [{\rm sin} i/(\mu_{\rm masy}d_{kpc})]^{4/3}
\eqno (11)
$$
then the former dominates (case A1) and the standoff angle (4) increases by  $x_i^{-1/2}$
(or the density inferred from a measured $\theta_a$ increases by $1/x_i$).
The minimum pre-ionization fraction satisfying equation (11) is 
$$
x_i=351 ({\dot E}_{34}/\theta_a)^{1/2} ({\rm sin} i/\mu_{\rm masy})^{3/2} d_{kpc}^{-2}
\eqno (12)
$$
where the observed standoff angle $\theta_a$ is in arcseconds, and sensible values obtain
for the strong case A (small ${\dot E}_{34}$, large $\mu_{\rm masy}$) limit. 
In turn this gives the maximum apex flux for case A1, which
is increased from equation (10) by $(1-x_i)/x_i$.

	In contrast, if the preshock ionization
fraction is sufficiently low, then post-shock ionization can dominate and 
from equation (4), replacing $n_H$ with $0.3\, r_a n_H/l$, we find (case A2)
$$
\theta_a=1.6^{\prime\prime} [ {\dot E}_{34} {\rm sin} i/(n^2 \mu_{\rm masy}d_{kpc}^4)]^{1/3}
\eqno(13)
$$
and
$$
f_{{\rm H}\alpha} =2.4 \times 10^{-4} f_{\rm HI}
\left ({ {{\dot E}_{34}^4 {\rm sin}\, i} \over {n_H^2 \mu_{\rm masy} d_{kpc}^{13}}}\right )^{1/6}
{\rm cm^{-2}\,s^{-1}}
\eqno (14)
$$
where the prefactors are doubtless somewhat sensitive to the details of the postshock
ionization, which can in principle be measured from flux profile at the
limb of well-resolved bow shocks. Notice that in case A2 $f_{{\rm H}\alpha}$ does depend
on the total upstream H density $n_H$.

\subsection{Survey Target Selection}

	From these expressions we see that a bow shock should have large
$\theta_a \propto {\dot E}^{1/2}/d^2$ to be resolvable and large 
$f_{{\rm H}\alpha} \propto {\dot E}/d^{7/2}$ to be bright. Thus, nearby, powerful
pulsars should dominate the bow shock sample. The most extensive uniform
compilation of such pulsars comes from the {\it Fermi} LAT, where in the
second pulsar catalog \citep[][hereafter 2PC]{2PC} 117 $\gamma$-ray pulsars are presented.
This represents an approximately flux limited sky survey. In this catalog, it
was noted that the pulsar $\gamma$-ray luminosity scales, heuristically, as 
$$
L_\gamma \approx (10^{33} {\rm erg\, s^{-1}} {\dot E})^{1/2}, \qquad  {\dot E} > 10^{33}  {\rm erg\, s^{-1}},
\eqno(15)
$$
so the flux at Earth scales as $f_\gamma \propto  {\dot E}^{1/2}/d^2$.
Interestingly, this means that (for $\mu$, $f_{\rm HI}$ etc. fixed) that
we have an approximate scaling $f_{{\rm H}\alpha} \sim f_\gamma^2$. Thus
bright $\gamma$-ray pulsars have the potential to show bright, resolvable
H$\alpha$ bow shocks.

	In view of the above, it is not surprising that most of the bow-shock
producing pulsars, both young and millisecond pulsars (MSP) are LAT-detected.
The exceptions are PSR J2225+6535 and J1856$-$3754, both low ${\dot E}$, high $v$
(case A) objects. PSR J1856$-$3754 is also very nearby and, lacking any 
non-thermal pulse detection, may be beamed away from Earth.
Thus, in searching for new pulsar H$\alpha$ bow shocks we focused on the 
LAT pulsars from 2PC and on nearby, energetic pulsars newly discovered in the direction
of {\it Fermi} LAT sources \citep[eg.][]{ret12,plet13}.

\begin{figure}[ht!!]
\vskip 8.0truecm
\includegraphics{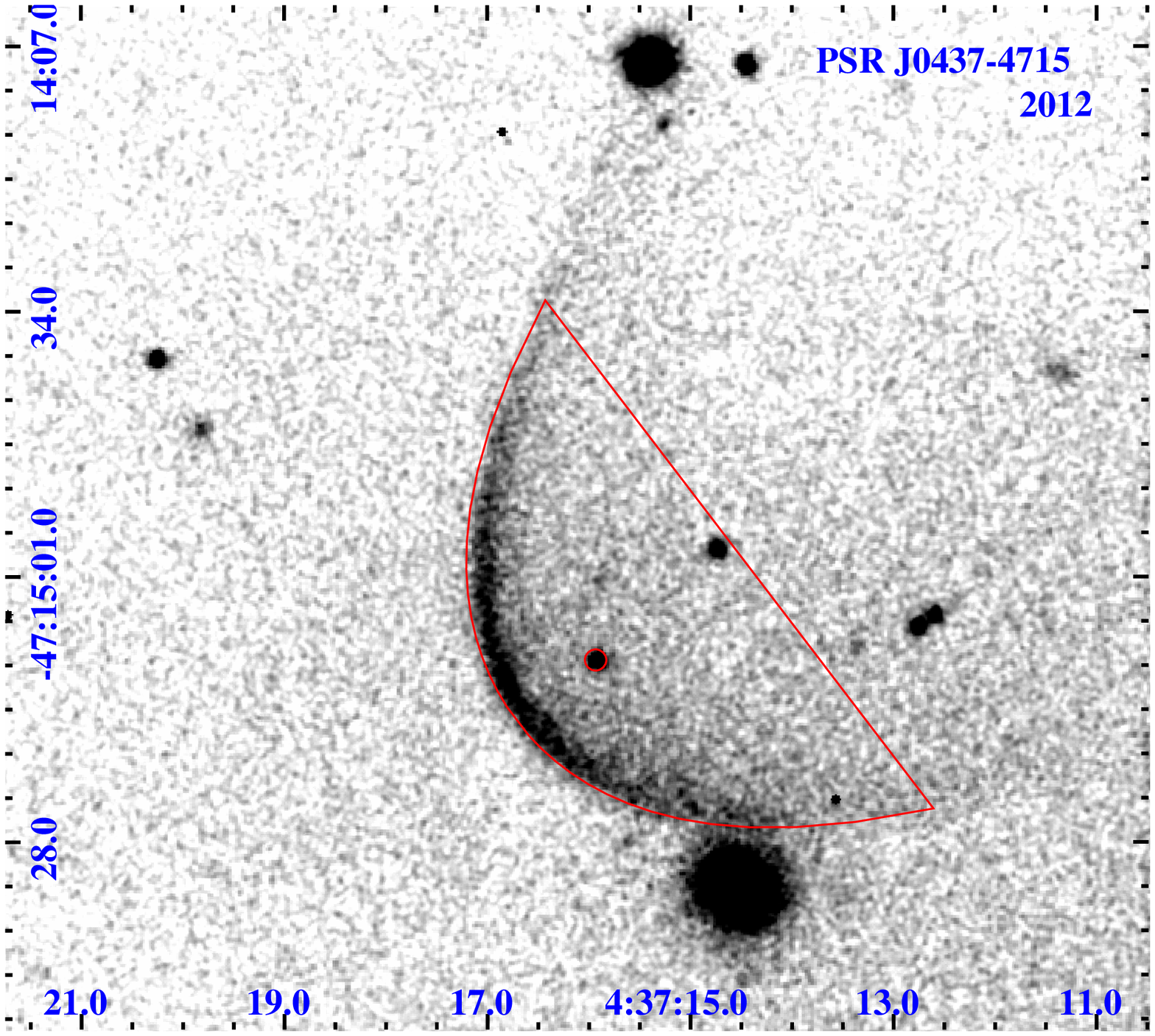}
\includegraphics{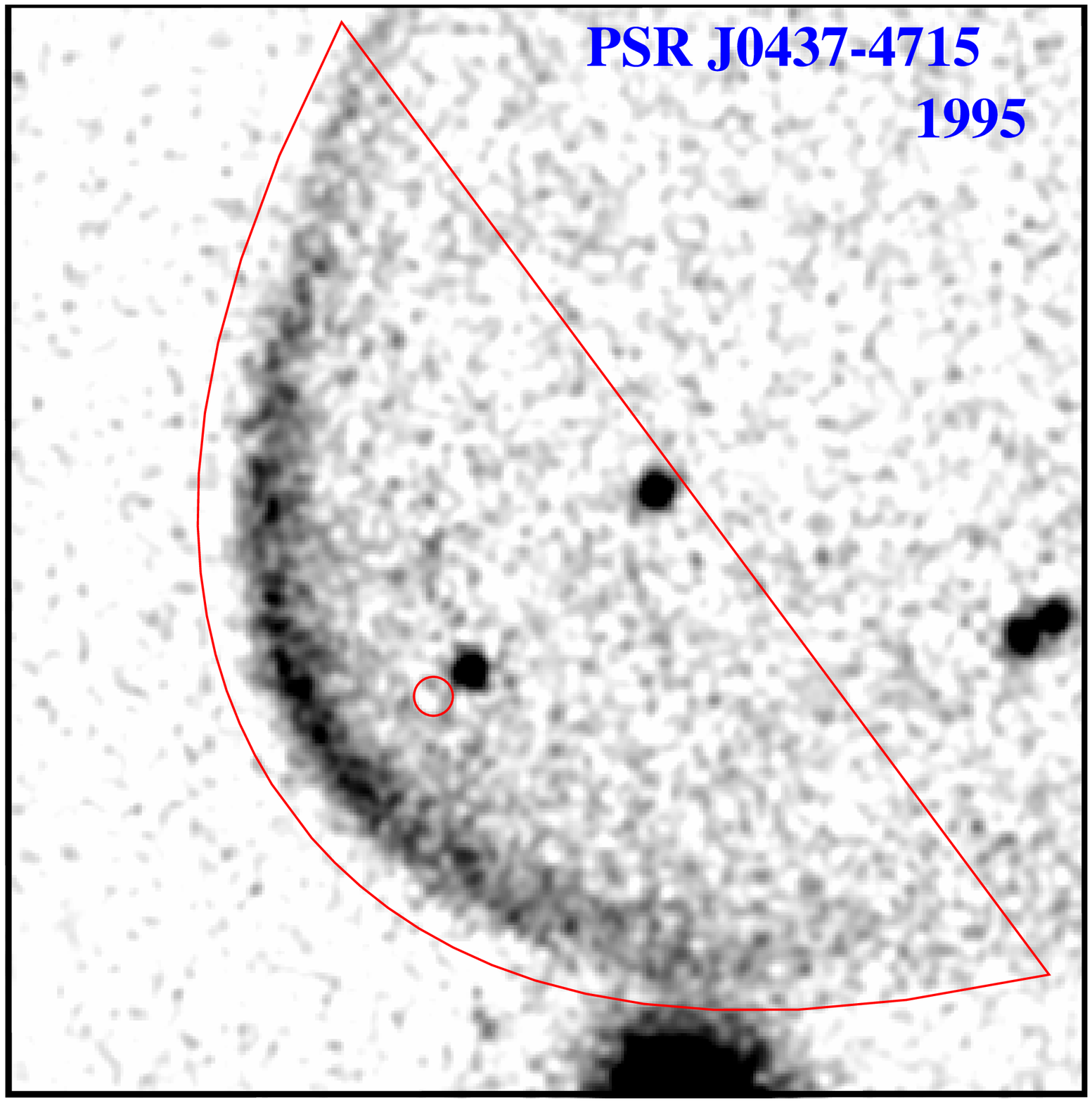}
\begin{center}
\caption{\label{J0437} $3\times$600\,s W012 SOI image of PSR J0437$-$4715. The line
delimits the apex region (eqs. 4-6).  The inset shows a 1995 H$\alpha$ image 
(Andy Fruchter, http://www.stsci.edu/$\sim$fruchter/nebula),
with the 2012 apex region and pulsar location indicated, for comparison. The 
visible pulsar companion and shock limb shift by $2.4^{\prime\prime}$, as expected.
}
\vskip -0.35truecm
\end{center}
\end{figure}

\section{Observations}

	Our pulsar H$\alpha$ campaign was pursued during a series of WIYN
and SOAR observations, allocated through the joint {\it Fermi}/NOAO program
to study the counterparts of LAT pulsar, blazar and unidentified sources. The
PWN data reported here came from portions of five observing runs, covering 
both the Northern and Southern hemispheres. For all H$\alpha$ images we used the WIYN 
W012 ($\lambda= 6566$\AA$,\,\Delta \lambda_{FWHM}= 16$\AA) narrow band H$\alpha$
filter, kindly loaned to SOAR for the two southern campaigns. To help distinguish
poorly-resolved H$\alpha$ sources near the pulsars from stars, we also obtained,
when possible, matching continuum observations using the W014 
($\lambda= 6562$\AA,$\,\Delta \lambda_{FWHM}= 378$\AA) wide H$\alpha$ filter
or an SDSS r$^\prime$ ($\lambda= 6163$\AA,$\,\Delta \lambda_{FWHM}= 1518$\AA)
broad band filter.

\subsection{WIYN Campaigns}

	In the North, we used the WIYN 3.6m telescope and the MiniMo camera,
with observations occurring on 2009 March 24-26, 2011 September 26-27
and 2012 February 17-18. The MiMo camera is a two chip mosaic, with 0.14$^{\prime\prime}$
pixels covering a 10$^\prime$ Field of View (FOV), with a 7$^{\prime\prime}$
gap.  The pulsar H$\alpha$ survey was the `poor seeing'
portion of the program and so generally images were taken under 0.9-1.4$^{\prime\prime}$
imaging. Despite the imaging limitations, we were able to maintain relative photometry to 
$\sim 10$\%, and substantial portions of the 2011 and 2012 runs
were near-photometric. 

In total 52 unique LAT pulsars were observed at WIYN (9 in 2009, 33 in 2011, 21 in 2012,
some re-observed). Typical initial exposures were 300s or 600s, objects showing promising
structure near the pulsar position received a second exposure.
Unfortunately, the relatively long MiMo read-out time (182s for a full frame) 
made short exposures inefficient and so continuum frames (30-60s)
were only obtained for a few pulsars. We observed the compact Balmer-dominated
planetary nebula M1-5=PN G184.0-0.21 \citep{wright05} to establish our flux scale
at $4.8 \times 10^{-5} {\rm H\alpha\, cm^{-2} s^{-1}/DN}$.

\subsection{SOAR Campaigns}

	We observed with the SOAR Optical Imager (SOI) on the 4.2m SOAR telescope
on March 21-23, 2012. SOI has 0.078$^{\prime\prime}$ native pixels in a 2 CCD mosaic
covering a 5.3$^\prime$ FOV with a 7.8$^{\prime\prime}$ gap. As we did not experience 
exceptionally good seeing, we ran binned $2\times2$, for a 0.15$^{\prime\prime}$ pixel
scale and an excellent read-out time of 11\,s. Again the best seeing during the run
was allocated to a different component of the program (BL Lac imaging) and 
pulsar H$\alpha$ images had typical 0.8-1.2$^{\prime\prime}$ delivered image quality.
The loss of mirror temperature control during the second half of the run adversely affected
image quality.  We observed 29 pulsar targets during this run. H$\alpha$ flux calibrations 
were generated from exposures of PN G232.8-04.7 \citep{dh97}, giving 
$5.9 \times 10^{-5} {\rm H\alpha\, cm^{-2} s^{-1}/DN}$. 60\,s W014 frames were obtained
for most targets to discriminate against continuum structures.

	We returned to SOAR on August 9-12, 2013, but for this run, the need for 
interleaved spectroscopy plus limited resources for instrument changes led us to
use the Goodman High-Throughput Spectrograph (GHTS) for the imaging. Although GHTS has
0.15$^{\prime\prime}$ pixels covering a 7.2$^\prime$ diameter FOV, the long
96s read-down led us to sub-frame to a 
$4.25^\prime \times 6.5^\prime$ field. In addition, poor seeing caused us to bin
$2\times2$ after the first night, resulting in a modest 15\,s readout. The seeing
was highly variable during the run, with brief periods at $\sim 1^{\prime\prime}$,
but typical seeing of $2^{\prime\prime}$, and periods over 5$^{\prime\prime}$.
The last night was largely lost to weather. Unfortunately, GHTS suffered electronic
pick-up during this run, leading to a variable baseline modulation of $\sim 10$DN
beyond the nominal read-out. Because of the very narrow W012 band-pass, sky count-rates
were low and this modulation dominated the background, limiting our sensitivity to
low surface brightness structures. Exposures were thus 600\,s in the W012 filter.
Calibration observations of the planetary nebula BoBn 1 \citep{wright05} gave
throughput estimates. However cross-comparison with fields common to the
SOI and MiMo runs indicated that the flux scale from this calibrator was high
by 20\%. Our adopted flux conversion after this correction was 
$5.1 \times 10^{-5} {\rm H\alpha\, cm^{-2} s^{-1}/DN}$, and we estimate the uncertainty
in the relative (and absolute) calibrations for the three systems as $\sim 10\%$,
both due to imperfect instrumental calibration and imperfect monitoring of transparency.
For this run, we elected to obtain SDSS $r^\prime$ continuum frames (typically 180\,s
exposure), which gave us adequate sensitivity to contaminating stars. This
also allowed modest sensitivity to stellar companions of binary LAT MSP, despite
the very poor imaging conditions. H$\alpha$ limits were obtained for 20 LAT pulsars
in this run.

\begin{figure*}[t!!]
\vskip 7.3truecm
\includegraphics{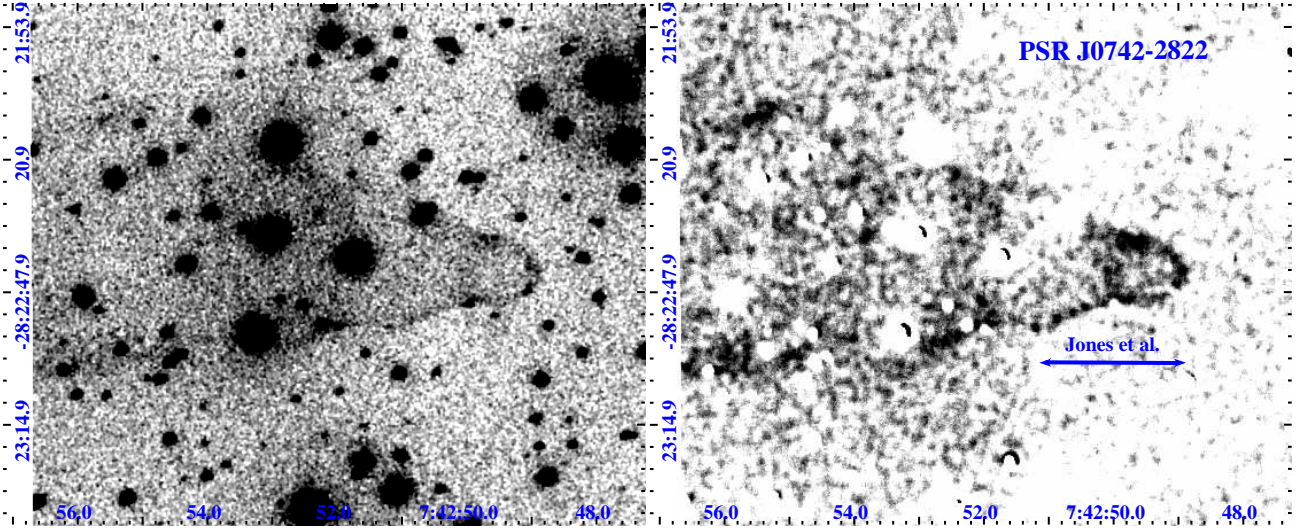}
\begin{center}
\caption{\label{J0742} Left: a median-filtered $3\times 600$\, W012 SOI image of 
PSR J0742$-$2822, smoothed with a $0.45^{\prime\prime}$ Gaussian.
The right panel shows an image with a scaled continuum (W014) image 
subtracted and $0.9^{\prime\prime}$ top-hat smoothing. The arrow indicates extent of the previous nebula detection.
}
\vskip -0.5truecm
\end{center}
\end{figure*}

\section{Measurement and Results}

	The data were subject to standard IRAF calibrations, with image bias subtraction and
correction using dome and sky flats. World coordinate systems were applied to the 
the frames, as appropriate, and the images were combined with a median filter, clipping
cosmic ray events. 

\subsection{New Images of Known Bow Shocks}

	To ensure that our survey sensitivity was sufficient to routinely detect 
pulsar bow shocks, we imaged several known objects. We observed J0437$-$4715 (March 22, 2012,
with SOI), where it was obvious in a single 300\,s exposure. In Figure 1 we show
a combination of $1800$\,s exposure under $\sim 1.2^{\prime\prime}$ imaging. The region shows our
`apex' zone from equations (5) and (6); these provide a very good match to the apex shape
with an effective standoff angle $\theta_a=9.3^{\prime\prime}$. In the inset we show this
2012 epoch shock limb and pulsar position on a 1995 image, showing the motion of the 
companion and bow shock apex over 17 years. 

	\citet{bell95} give the H$\alpha$ flux of this nebula as $2.5 \times 10^{-3} 
{\rm cm^{-2} s^{-1}}$, and this has been widely quoted. We have measured the H$\alpha$ flux
in the apex zone (expanded slightly to account for the image FWHM) and find 
$6.7 \pm 0.7 \times 10^{-3} {\rm cm^{-2} s^{-1}}$. We also took $2\times 600$\,s exposure
with the GHTS on August 17, 2013. Although the seeing was very poor we were able to confirm
the large flux, obtaining $\approx 6.9 \pm 1.0 \times 10^{-3} {\rm cm^{-2} s^{-1}}$, where the uncertainty
is dominated by the choice of background for this large diffuse nebula. The {\it full} nebula
contains $\sim 60\%$ more flux. Thus we find the the H$\alpha$ bow shock flux of PSR J0437$-$4715 is
$2.7-4.3\times$ larger (depending on region) than previously reported; this will be important 
to our discussion below.

	To check sensitivity against the faintest young radio pulsar bow shock reported
in the literature, we also observed PSR J0742$-$2822 with SOI. Here the shock was
clear in a single 600\,s exposure. Figure 2 shows a $3\times 600$\,s median stack.
Guided by the 6500\,s NTT discovery image of \citet{jet02}, we centered the pulsar
on one of the array chips; we see a rather similar `keyhole' shape at the apex. 
However, our combined data show that the nebula extends much further than previously reported.
\citet{jet02} suggest that the nebula closes off some $45^{\prime\prime}$ behind the pulsar
but we see the wedge of H$\alpha$ emission extending off of our frame, over $2.5\times$ further.
This tail shows multiple swellings reminiscent of the bubbles of PSR J2225+6535/Guitar nebula.
Doubtless faint emission extends beyond our image, so we have not captured the full size of this
nebula.  These data confirm that the very narrow bandpass of the W012 filter provides
excellent contrast for the H$\alpha$ emission; although the image is count limited, we
obtain S/N comparable to that of \citet{jet02} in $\sim 1/5$ the exposure time.

\begin{figure}[ht!!]
\vskip 7.3truecm
\includegraphics{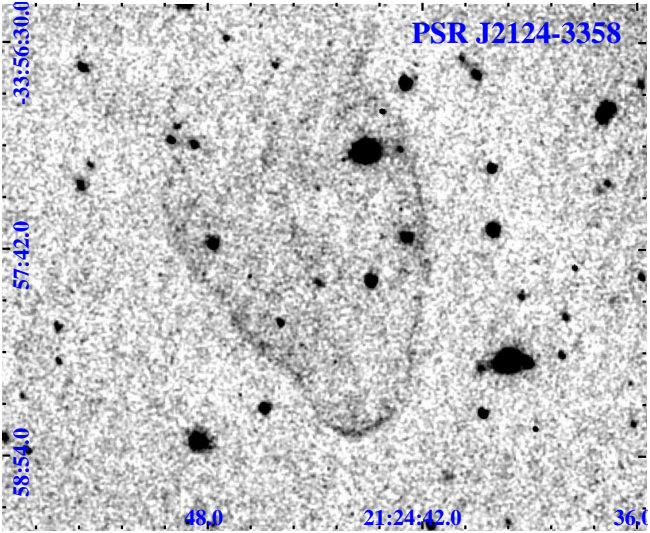}
\begin{center}
\caption{\label{J2124} 600\,s W012 GHTS image of PSR J2124$-$3358, smoothed with a
$0.9^{\prime\prime}$ Gaussian. The bow shock narrows $70^{\prime\prime}$
behind the pulsar but appears to continue faintly to larger $r/r_a$. The background
diagonal striations are pick-up noise during read-out.
}
\vskip -0.55truecm
\end{center}
\end{figure}

	Similarly we observed the faintest known MSP bow shock, that discovered by
\citet{ganet02} around PSR J2124$-$3358 in a 4800\,s NTT image stack. We 
exposed a single 600\,s frame, using the GHTS on August 10, 2013, under poor ($\sim 2^{\prime\prime}$)
seeing and with the large pattern noise at readout. Nevertheless, this image (Figure 3)
shows more structure than the discovery data -- for example the nebula
narrows to the north of the pulsar, making the bulk of the body into a cavity.
Also there is no sign of the filament to the East of the PWN seen in the NTT data;
this may have been an optical artifact during that exposure.
However the `kink' structure on the east of the bow shock and the bright limb at the apex
are very well measured. Again, this good detection gives us confidence that we
would usually detect structures as bright as those previously reported in the literature.

\begin{figure*}[ht!]
\vskip 7.3truecm
\includegraphics{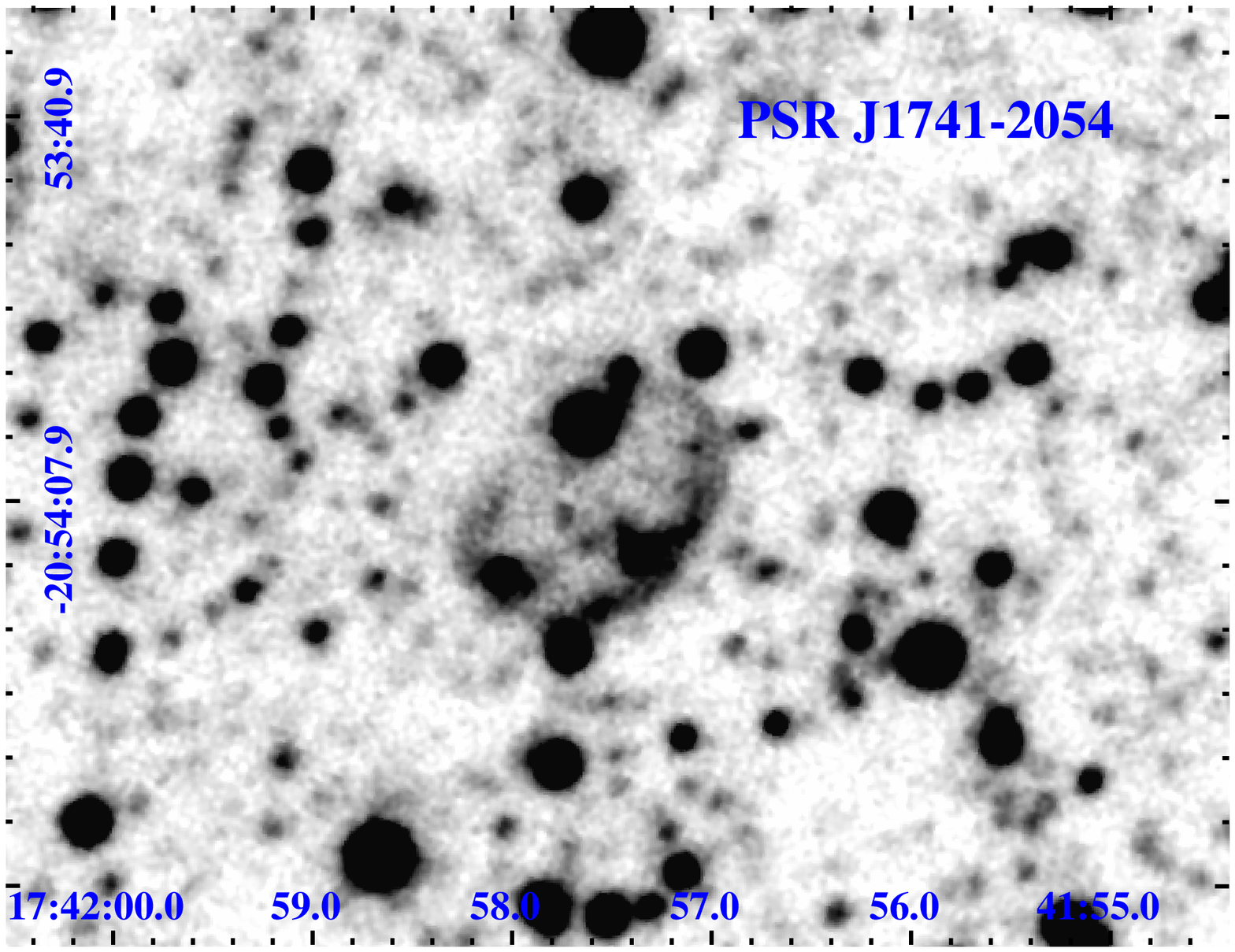}
\includegraphics{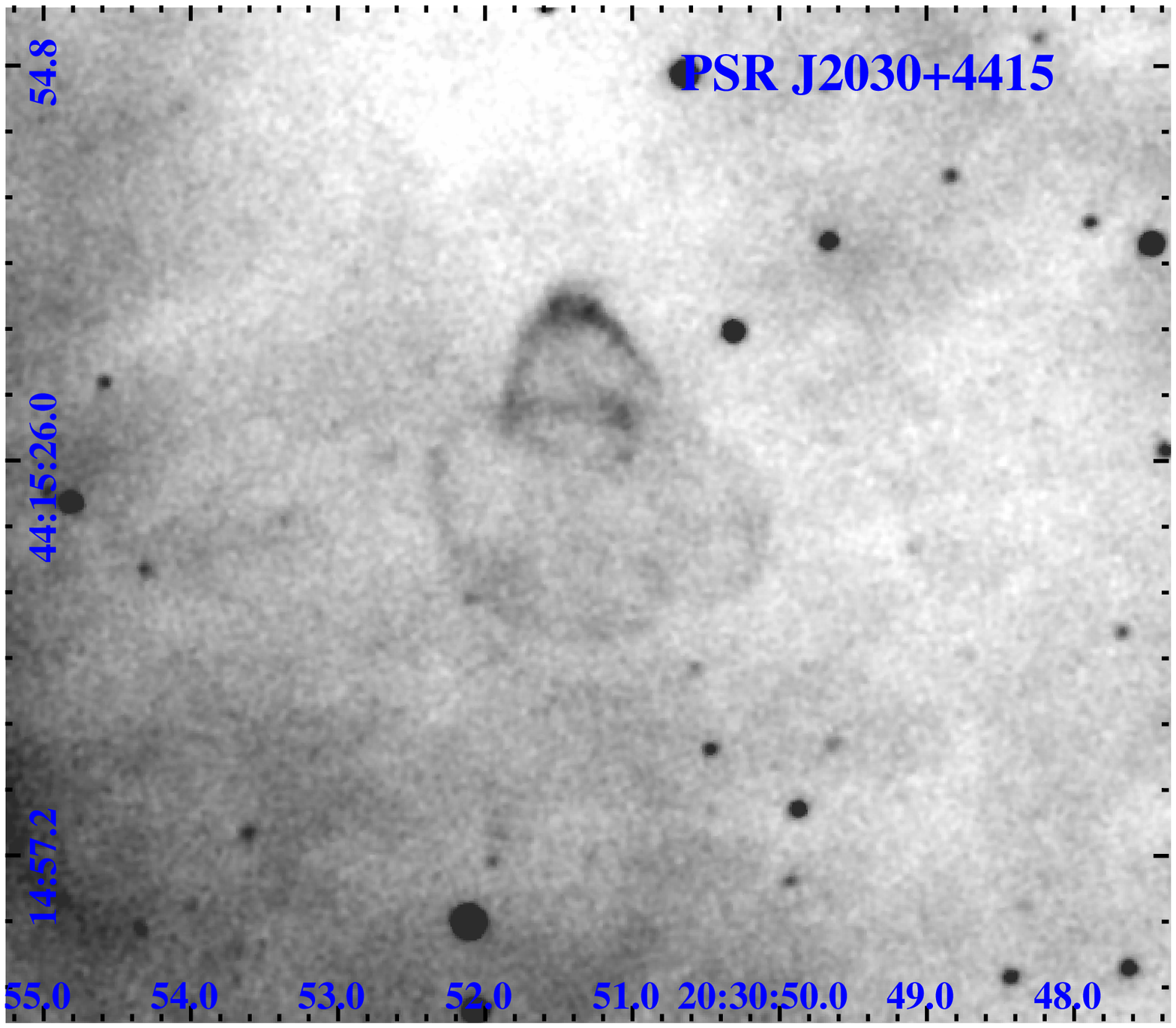}
\begin{center}
\caption{\label{J1741/J2030} Left: a median-filtered $1800$\,s W012 
MiMo image of PSR J1741$-$2054, smoothed with a $0.45^{\prime\prime}$ Gaussian \citep{ret10}.
Right: median-filtered $3\times 600$\, W012
MiMo image of LAT pulsar PSR J2030+4415, embedded in diffuse H$\alpha$. Faint emission appears
$\sim 2^{\prime\prime}$ ahead of the bow shock.
}
\vskip -0.25truecm
\end{center}
\end{figure*}

\subsection{New Pulsar H$\alpha$ Structures}

	We briefly describe the morphology and flux of the newly detected H$\alpha$
structures to facilitate comparison with previously known bow shocks and our survey 
upper limits. Detailed modeling and multiwavelength analysis is deferred to
future publications. We discuss the novel pre-ionization structures seen
for PSR J1509$-$5850 and J2030+4415 in \S4.4

	The bow shock of PSR J1741$-$2054 has been described in \citet{ret10}. It is
quite bright, but lies in a very crowded Galactic field. This nebula hosts a bright
X-ray PWN, which trails off to the NE along the H$\alpha$ minor axis. Perhaps
the most interesting aspect of this PWN is that the pulsar lies rather close to the forward shock
apex in projection and the apex curvature is small; it was argued that this indicates
an equatorially concentrated pulsar wind. As for PSR J2124$-$3358, this induces us to
use a $\theta_a \approx 2\times$ larger than the observed standoff to set the apex scale
and the effective apex H$\alpha$ flux in Table 1.

	PSR J2030+4415 is a 227\,ms, ${\dot E}_{34}=2.2$ pulsar found in a blind search of
an unidentified {\it Fermi} source \citep{plet12}. This pulsar has not been detected in the radio.
Figure 4 shows a $3\times$600\,s W012 frame taken September 26, 2011 with WIYN/MiMo, showing a very bright
compact H$\alpha$ bow shock coincident with the pulsar. The apex stand-off is estimated as
$\theta_a = 1^{\prime\prime}$, while the apex zone provides $2.4 \times 10^{-3} {\rm cm^{-2}s^{-1}}$.
After the apex, the PWN shows a closed bubble covering $\sim 15\times 25^{\prime\prime}$.
The apex appears partly superimposed on this bubble, so the pulsar velocity may have significant
inclination to the plane of the sky.
The region shows extensive diffuse H$\alpha$ emission, yet the sharpness of the nebula limb indicates
that the pulsar is embedded in largely neutral H.

	This pulsar lacks any parallax or DM distance estimates. However, we can get a
crude estimate for the distance by invoking the phenomenological $\gamma$-ray luminosity
law (equation 15) to find
$$
d_\gamma = (f_\Omega L_\gamma /4\pi F_\gamma)^{1/2} \approx 1.6{\rm kpc}(f_\Omega/F_{-11})^{1/2} {\dot E}_{34}^{1/4}
\eqno(16)
$$
where the LAT flux is $F_{-11} 10^{-11} {\rm erg\, cm^{-2}s^{-1}}$ and
the $\gamma$-ray beaming fraction is typically $f_\Omega \approx 1$, although it
can range from $\approx 0.1-3$ depending on the pulsar magnetic inclination $\alpha$
and viewing angle $\zeta$ \citep{wet09}.  With an observed $\gamma$-ray flux 
$F_{-11} = 5.8$ for PSR J2030+4415 (2PC) we infer a distance d=0.88\, kpc.

	One interesting feature of this bow shock is the faint diffuse H$\alpha$ seen ahead of
the apex shock limb.  This is likely photo-preionization; we have examined 
14\,ks of archival SWIFT XRT 
data and see a clear detection of the pulsar/PWN, with an unabsorbed 0.3-10\,keV flux of 
$1.4 \times 10^{-13} {\rm erg\, cm^{-2} s^{-1}}$, for a typical $\Gamma=2$ and 
$N_H \approx 3 \times 10^{21} {\rm cm^{-2}}$. While these values require confirmation from higher sensitivity
X-ray observation, this implies an X-ray luminosity 
$L_X \sim 1.6 \times 10^{31} d_{kpc}^2 {\rm erg\, s^{-1}}$ from the vicinity of the pulsar.

In Figure 5 we show H$\alpha$ structure associated with the young ($P=89$\, ms, $\tau=1.5\times 10^5$y)
energetic $5.2\times 10^{35} {\rm erg\, s^{-1}}$ radio/$\gamma$ pulsar J1509$-$5850. This pulsar
has a long $>8^\prime$ PWN trail observed in the radio \citep{nget10} and X-ray
\citep{ket08}. The X-ray flux of the PWN
is $\sim 2.4 \times 10^{-13} {\rm erg\, cm^{-2} s^{-1}}$, with $\sim 1/2$ of this flux from near the
pulsar position. In the right panel we overlay a portion of this X-ray trail from a 40\,ks
archival {\it CXO} image (OBSID 3513).  The brightest arc of the extended emission follows
the bow shock structure near the pulsar.  With a DM estimated distance of 2.6\,kpc, this 
apex region is relatively luminous with $L_X > 1 \times 10^{32} {\rm erg\, s^{-1}}$.
When smoothed on larger scales fainter X-ray emission is visible, extending some $8^\prime$ 
behind the pulsar.
An additional 380\,ks of {\it CXO} exposure is being collected on this pulsar, which should
allow an excellent study of the arcsec-scale X-ray PWN and its relation to the H$\alpha$ structure.

In contrast to J2030+4415, the H$\alpha$ emission is dominated by a spherical halo centered on the
pulsar, almost certainly due to X-ray excitation of the upstream medium. The most prominent
feature of the bow shock is the cavity in this halo, where the relativistic pulsar wind eliminates
the HI. However, this cavity is significantly edge-brightened, indicating that the upstream medium
shocks and that the HI surface brightness increases in this zone. The brightest limb patches (where the
bow shock flares well behind the pulsar) have a surface brightness peaking at $\sim 4 \times 10^{-5}
{\rm cm^{-2} s^{-1} arcsec^{-2}}$, although the average brightness along the limb is $\sim 10\times$ less.
We estimate a standoff scale $\theta_a=1.2^{\prime\prime}$ and a flux $1.4 \times 10^{-4}
{\rm cm^{-2} s^{-1}}$. This is embedded in a halo of radius $30^{\prime\prime}$ and
flux $6.6 \times 10^{-3} {\rm cm^{-2} s^{-1}}$. These structures have high statistical significance, 
but given the rather low H$\alpha$ surface brightness they are not visually striking; 
we outline their edges in the right panel to guide the eye.

\subsection{The H$\alpha$ Pulsar Bow Shocks}

	Scaling laws for the flux of H$\alpha$ bow shock emission have in the past attempted
to model the entire shock flux. This is difficult as the detailed limb shape and emissivity
seem to be strongly dependent on local variation in the ISM and possibly on instabilities in the
back-flow of the shocked relativistic pulsar wind (e.g. J0742$-$2822, J2030+4415
and J2225+6563). Some studies \citep[e.g.][]{crl93,cc02} suggest that with higher pulsar 
velocity the shock can support H$\alpha$ production with an increasingly oblique shock
so that the total H$\alpha$ flux scaled as $\sim v^3$. This is physically plausible, but 
certainly subject to the ISM variation and instabilities noted above. In contrast, when the 
pulsar apex region is resolved, the bow shock geometries are very similar. We 
choose here to concentrate on this region as it has the best hope of correlating with, 
and being a useful probe of, the pulsar properties.

\begin{figure*}[ht!!]
\vskip 8.5truecm
\includegraphics{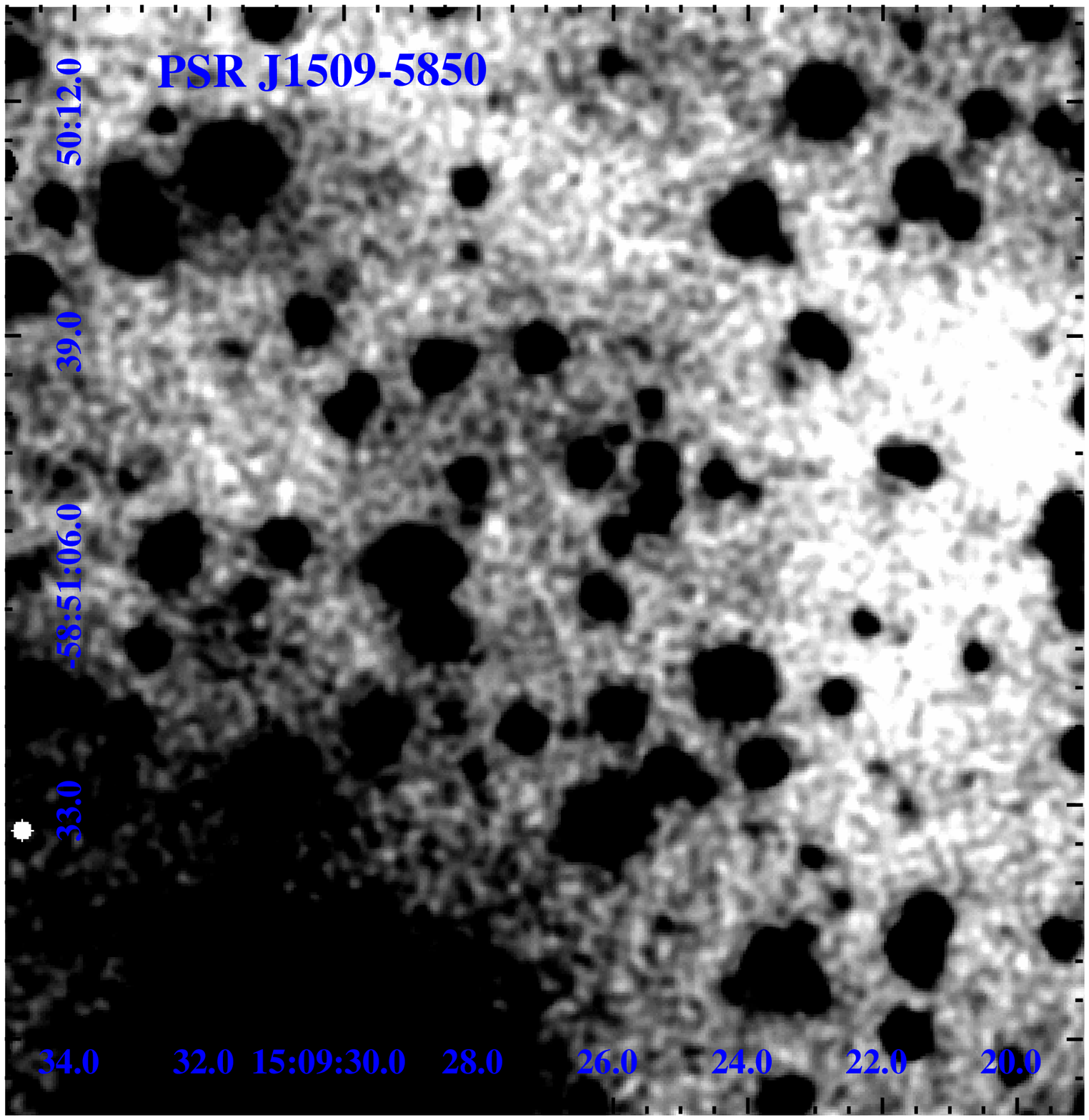}
\includegraphics{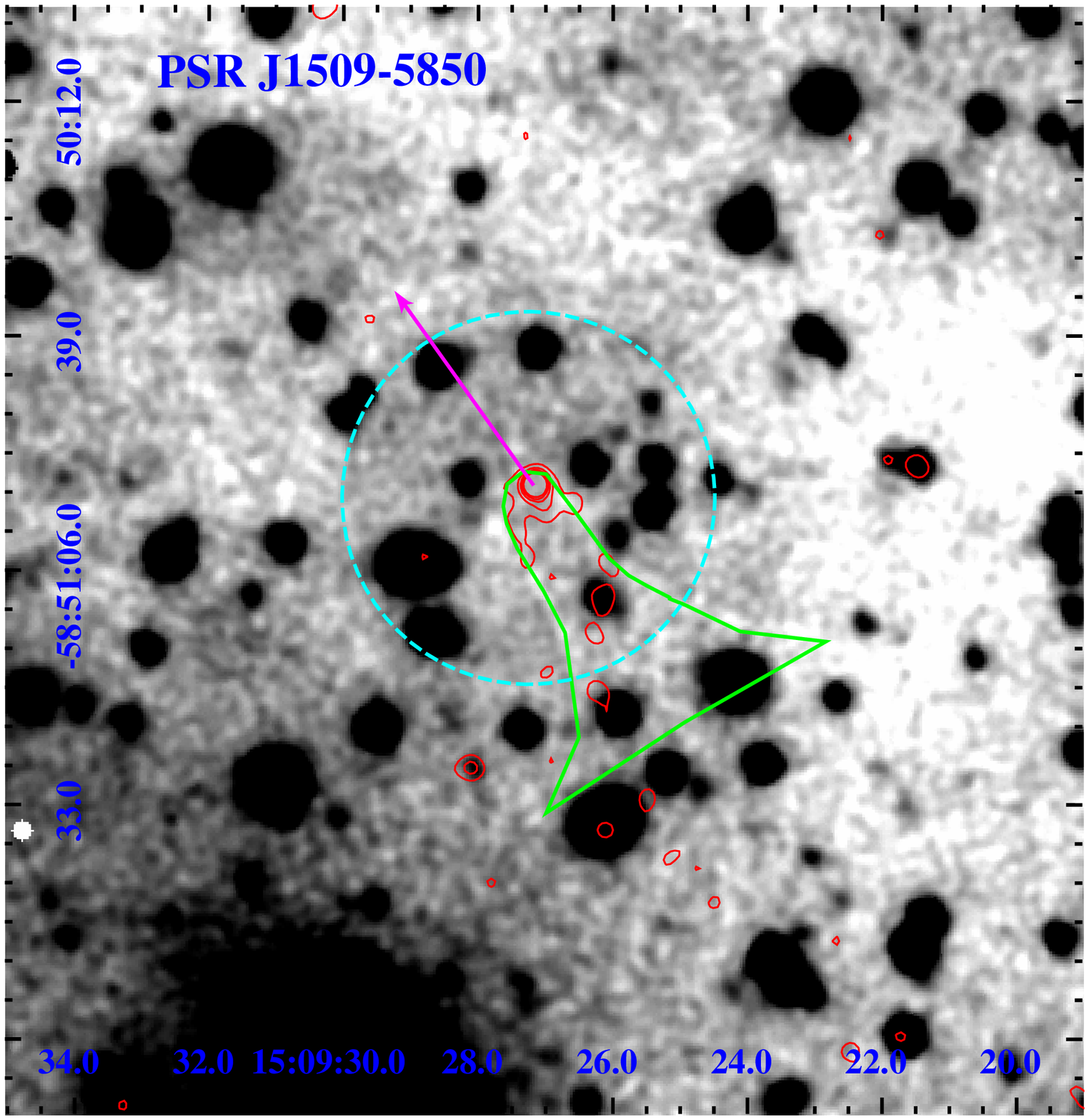}
\begin{center}
\caption{\label{J1509} Left: a $2\times 600$\, W012 GHTS image of PSR J1509$-$5850, smoothed 
with a $1.2^{\prime\prime}$ Gaussian. A faint H$\alpha$ halo surrounds the pulsar, with an 
edge-brightened bow-shock cavity. 
The stretch is hard, to bring out low surface brightness structure.
The right panel indicates the pulsar proper motion inferred from the X-ray trail. A dashed
circle shows the halo and a solid outline shows the bow shock cavity. Superimposed are 0.7-7\,keV
contours from an archival CXO image \citep[see][]{ket08}. The pulsar point source lies slightly
off the bow shock axis, but the extended X-rays outline the apex.
}
\vskip -0.35truecm
\end{center}
\end{figure*}

	To this end, we compile in Table 1 the observed properties for the pulsars with 
resolved H$\alpha$ emission at the apex. The spindown powers listed have been corrected for
the \citet{shk70} effect at the default distance.

\begin{deluxetable*}{lcrrrrrlrrrr|rrrr}[ht!!]
\tablecaption{\label{HaBS} Pulsars with H$\alpha$ Bow Shocks}
\tablehead{
\colhead{Pulsar}&\colhead{${\dot E}_{34}^a$}&\colhead{Lg$\tau$}&\colhead{d$^b$}&\colhead{$\mu_T$} 
&\colhead{$F_\gamma^c$}       &\colhead{$F_{x,NT}^c$}             &\colhead{$\theta_a$}&\colhead{$F_{a{\rm H}\alpha}$}
&$r/r_a$ &\colhead{$F_{T{\rm H}\alpha}$} &\colhead{$A_R$}
&\colhead{v$_\perp$}& \colhead{$n_{fit}$}&\colhead{$r_{apex}$}&\colhead{$\theta_I$\qquad}
\cr   
                &$\rm erg/s$              &y               &kpc        &mas/y  
&$10^{-11}$&$10^{-13}$&$^{\prime\prime}$  &$\gamma$/cm$^2$/s& &$\gamma$/cm$^2$/s &mag 
&km/s &${\rm cm^{-3}}$ & max &$^{\prime\prime}$
}
\startdata
J0437$-$4715&0.55&9.8&0.16 P&     141.3& 1.67&   7.9& 9.3&6.7E-3& 5  &1.1E-2&0.01& 107&0.21& 0.87 & 1.2\\
J0742$-$2822&19.0&5.2& 2.0 D&      29.0& 1.72&$<$0.2& 1.4&1.8E-4&$>$70&7.6E-3&0.87& 275&0.28&1.09&$<$0.2\\
J1509$-$5850&68.2&5.2& 2.6 D&          &12.70&   3.0& 1.2&1.4E-4& 30 &6.6E-3&3.51& -- &6.14& 0.98&   8.2 \\
J1741$-$2054&12.6&5.6&0.38 D&          &11.70&   2.0& 2.3&4.6E-3& 6  &7.6E-3&0.66& -- &1.44& 2.68& 0.9\\
J1856$-$3754&3.E-4&6.5& 0.16 P&   332.0&  -- &   0.0&0.85&3.E-5&$>$25&9.0E-5&0.06& 252&{\it .003}&{\it 29.6}& 0.4\\
           &     &  &        &        &     &      &    &     &     &      &  A1& 252&0.05&1.25& 0.4\\
J1959+2048  &21.9&9.5& 2.5 D&      30.4&  1.7&   0.7& 3.6&1.8E-3& 17 &5.6E-3&0.17& 360&{\it 0.02}&{\it  11.4}& 0.5\\
J2030+4415  &2.90&5.8& 0.9 G&          &  5.8&   2.8& 1.1&1.8E-3& 10 &9.3E-3&0.50& -- &2.69& 2.05&2.4\\
J2124$-$3358&0.68&9.8& 0.30 P&      52.7&  3.7&   0.8& 5.0&5.3E-4& 25 &5.6E-3&0.12& 75 &0.47& 0.14& 0.3\\
J2225+6535  &0.16&6.1&1.86 D&      182.0&  --&   0.0&0.12&3.6E-5&100 &4.6E-3&0.57&1610&{\it 0.01}&{\it  240.}& 0.0\\
  && &1.00\,\,&           &    &      &    &      &    &      &  A2& 866&1.43& 1.42& 0.0\\
\enddata
\tablenotetext{}{Tabulated quantities, in order: 
Pulsar name, spindown power, characteristic age, distance/method, transverse proper motion,
GeV flux, PSR/PWN head X-ray non-thermal flux, estimated standoff angular scale, apex H$\alpha$ flux,
full nebula to apex size ratio, full nebula H$\alpha$ flux, estimated extinction, perpendicular velocity,
fit total upstream density, ratio of observed apex flux to model flux, characteristic ionization angular scale.
See text for details.}
\tablenotetext{a}{Computed at the nominal distance $d$. All young pulsars are assumed to have $M=1.4\,M_\odot/
I_{45}=1.3$. For PSR J0437$-$4715 \citet{veret08} measure $M=1.76\pm0.2$, for J1959+2048 \citet{vKK11} measure
$M=2.40\pm0.12$; other MSP are assigned $M=1.8\,M_\odot/I_{45}=1.9$.
The spindown power for PSR J1856$-$3754 comes from X-ray timing \citep{kk08}.
}
\tablenotetext{b}{ Distance from D=dispersion measure, G=$\gamma$-ray flux, P=parallax.}
\tablenotetext{c}{X-ray and $\gamma$-ray fluxes in ${\rm erg\,cm^{-2}s^{-1}}$.
Values are from 2PC, except for J1741, J2030 (X-ray re-measured) and J1856, J2225 (undetected in $\gamma$-ray).}
\end{deluxetable*}
\bigskip

	For six of these pulsars we measured the bow shock apex scales and fluxes from our calibrated 
images (Figures 1-5), and estimated the full extent of the H$\alpha$ emission. It is
difficult to assign flux errors, as these are dominated by uncertainty in the definition
of the nebula boundary, exclusion of faint continuum sources and, for the larger nebulae, removal
of the spatially variable diffuse background. We estimate $\sim 20$\% for these systematic errors.
For J1741$-$2054 and J2124$-$3358, the flattened apex indicates an equatorially concentrated
pulsar wind \citep{ret10} -- for these we use an apex zone scaled to an effective standoff 
$\theta_a \approx 2\times$ the actual distance from the pulsar to the shock limb; this better 
fits the curvature of the apex and estimates the ISM cross section.

	We do not have new images for three of the bow shocks. 
For J1856$-$3754 we scale from the $2\times 10^{-5}{\rm H\alpha\,cm^{-2}s^{-1}}$ (within a 
perpendicular distance of $2^{\prime\prime}$) given by \citet{vKK01}, the estimate of the 
full bow shock flux as $\sim 3\times$ that of the apex is rough as it becomes faint downstream.
For J1959+2048, in archival images we find for the full nebula a somewhat smaller flux than 
the $7.3\times 10^{-3} {\rm cm^{-2}s^{-1}}$ given by \citet{kh88}. The tabulated apex flux
comes from a new calibrated integral field observation of the nebula (integrating over the
H$\alpha$ line); we cross check this flux against observations of J1741$-$2054 from the 
same night. These data will be described elsewhere (Romani et al. 2014, in prep.).
For J2225+6535 we have used the calibrated archival {\it Hubble Space Telescope} WFPC2 images 
described by \citet{cc02}. Like these authors, we find that the apex varies significantly between the 
two epochs, with $\theta_a \approx 0.10^{\prime\prime}$ in 1994 and $\approx 0.14^{\prime\prime}$ in
2001; the surface brightness is $\sim$constant and the apex flux is $\sim 3\times$ brighter in 2001.
Table 1 lists the average of these epochs. The `total' length and flux are for the guitar head
as this seems to correspond most closely to the other bow shocks. Including the guitar body,
the full nebula extends $82^{\prime\prime}$ (680$r_a$) and delivers $8.6\times 10^{-3}
{\rm cm^{-2}s^{-1}}$ of H$\alpha$. The low surface brightness of this structure makes
an accurate flux measurement difficult.

	The left sector of Table 1 contains measured quantities.  
Our goal here is to compare these observations with the estimates of \S2. The right sector
of the table contains a few inferred quantities, starting with the perpendicular space velocity for
the estimated distance. When the distance estimate is from parallax (P), this is reasonably secure. The next
column gives $n_{fit}=n_H/{\rm sin}^2 i$, as inferred from equation (4); when we lack a proper motion
we assume $v_\perp = 100{\rm km\,s^{-1}}$. $r_{apex}$ gives the ratio of observed apex flux to the
maximum for a fully neutral medium given by Equation (10) or Equation (14); again 
$v_\perp = 100{\rm km\,s^{-1}}$ is assumed if we have no information. 

	Most apex values agree well with the model, especially considering the measurement difficulties
and approximations involved in generating our model scaling laws.
Three pulsars stand out as glaring exceptions. For PSRs J1856$-$3754 (in the total ionization limit),
J1959+2048 and J2225+6535, the model ISM density $n_{fit}$ is very small and the flux ratio $r_{apex}$ 
very large. In particular the J2225+6535 apex is $240\times$ brighter than expected. Two factors contribute
to this mismatch. The assumption of complete post-shock ionization cannot be true for the
low luminosity non-LAT pulsars J1856$-$3754 and J2225+6535, since in this limit they have $l/r_a=$
345 and 280 (at 1.86\,kpc), respectively. Thus the gas is only partly ionized before
it reaches the contact discontinuity;
Equation (11) determines whether ambient ionization
or post shock ionization dominate. For PSR J1856$-$3754 we find that the upstream ionization will
maintain case A1 if $x_i > 0.043$. This would lead to a fit upstream density of
$0.052 {\rm cm^{-3}}$ and a model apex flux of $2.4\times 10^{-5}{\rm cm^{-2}s^{-1}}$ ($r_{apex}$=1.25). 
Slightly higher preshock ionization (likely, considering this pulsar's  modest, but non-negligible,
ionizing X-ray flux; \S4.4) increases the allowed density and lowers the apex flux. If the density
is too high, with low $x_i$, then case A2 applies. Then equations 13 and 14 give a density  
$n=0.10 {\rm cm^{-3}}$
and apex flux $f_{\rm H\alpha}= 4.6 \times 10^{-5} {\rm cm^{-2}s^{-1}}$ ($r_{apex}$=0.64).
This pulsar is evidently very close to the border between these two cases, as the predicted fluxes
bracket the observed value.  For PSR J2225+6535,
the critical pre-ionization fraction is 0.1 at $d=1.86$\,kpc. For higher densities, the model
gives $n= 0.12 {\rm cm^{-3}}$, while underpredicting the H$\alpha$ flux by $7.4\times$ (case A1)
or $n= 0.42 {\rm cm^{-3}}$, underpredicting $3.6\times$ (case A2), for lower preionization densities.

	The third object in question is J1959+2048, an energetic LAT pulsar. With $l/r_a=0.06$ this
should reach full ionization in the post-shock flow. However J1959+2048 (and, at present, J2225+6535) 
have only DM distance estimates. With the strong $d_{kpc}^{-2}$ and $d_{kpc}^{-7/2}$ dependencies
in Equations (4) and (10), even modest distance revisions can greatly increase the expected flux.
For J1959+2048, matching $r_{apex}$ via equation (10) gives $d=1.25$\,kpc, but since the Shklovskii 
correction is substantially reduced, the fit distance is actually $d=1.4\,$kpc. At this distance the velocity 
$v_\perp=202\,$km/s and upstream total density $n_{fit} = 0.28\,{\rm cm^{-3}}$ are quite plausible.
In this calculation we have used the large $I_{45}=2.9$ implied by equation (1); smaller moments
of inertia require even smaller distances. For comparison, \citet{arc02} found a $\sim 1.2$\,kpc 
distance from bow shock modeling of this pulsar.

	For J2225+6535, while incomplete post-shock ionization increases the expected flux, the
model at $d=1.86$\,kpc is still at least  $3.6\times$ fainter than observed. In table 1 we show
that the case A2 calculation for 1\,kpc gives an expected flux 70\% of that seen. In fact the
values match at 0.8\, kpc. However, for case A2 the flux depends on $n_H$ and thus the measured 
standoff and $\theta_a$. Given the difficulty of defining the apex region and 
its flux, even with {\it HST}, we should only infer $d\approx 1$\,kpc for this pulsar.
Similarly the  $n_{fit} = 1.4\,{\rm cm^{-3}}$ density estimate is very sensitive to the
poorly resolved (and apparently variable) standoff angle.
A high precision parallax distance will be particularly interesting as it will drive improved 
understanding of the emission of this peculiar shock.

	PSRs J0437$-$4715, J0742$-$2822 and J1509$-$5850 have $r_{apex} \approx 1$. The small value for 
J2124$-$3358 implies a relatively low neutral fraction in the upstream medium. A more detailed analysis
of the shock apex, including anisotropy of the pulsar wind is needed for a more precise estimate of the neutral
fraction. For the last two pulsars very modest $<25\%$ adjustments to the DM distances bring them into line.
For J1741$-$2054, the apex flux suggests $d=0.29$\,kpc, which seems plausible  given the very small DM.  For J2030+4415, 
8ur only distance estimate is the very crude 0.88\,kpc from the observed $\gamma$-ray flux. The H$\alpha$ apex 
flux implies $d=0.72$\,kpc. Thus the expected incomplete postshock ionization for low ${\dot E}$ pulsars together with
adjustment to the relatively uncertain DM distances allows us to model the observed $r_{apex}$ for
{\it all} bow shock pulsars, certainly to within the accuracy of our apex measurements.
This success, together with the strong $d$ dependence implies that the apex flux provides
a new, sensitive pulsar distance estimator. We discuss this further below.

\subsection{Precursor Ionization Halos?}

	The diffuse H$\alpha$ emission centered on the pulsars in our images
of J2030+4415 and, especially, J1509$-$5850 suggest a new PWN component, an ionization
pre-cursor. Such emission is not unanticipated: \citet{blaet95} proposed that accreting isolated
neutron stars could ionize cometary HII regions and \citet{vKK01} applied these ideas to
J1856$-$3754. In the latter paper a numerical model was run which, while not matching the
edge-brightened shock structure of J1856$-$3754, predicted an H$\alpha$ halo centered on the
pulsar with a PWN-evacuated trail (their Figure 4) which bears some similarity to our
image of PSR J1509$-$5850 (Figure 5).
This suggests that pre-shock ionization does produced the H$\alpha$ halo. We would
like to understand its origin and its dominance for J1509$-$5850, in contrast to
other pulsar bow shocks. 

	J1509$-$5850 has substantially the largest spin-down luminosity in our bow shock
set, suggesting that the non-thermal flux dominates the pre-ionization. 
A useful quantity is ionization standoff angle
$$
\theta_I \approx \int {\alpha_{\rm ion}(\nu) {\dot N}_{\rm ion}(\nu){\rm d}\nu}/(\pi v d)
\eqno (17)
$$
where $\alpha_{\rm ion} = 2.0 \times 10^{-23}E_{\rm keV}^{-3}{\rm cm^2}$ is the H photo-ionization 
cross section above 13.6\,eV and ${\dot N}_{\rm ion}(\nu)$ is the spectrum of the apex region, 
including the pulsar itself. We will estimate ${\dot N}_{\rm ion}$ from X-ray observations of 
the pulsar/PWN apex. Full surface temperatures of young neutron stars are $kT \sim 50$\,eV while 
millisecond pulsar thermal emission is often dominated by a smaller, hotter zone. In both cases
the ionization is dominated by photons well down on the Rayleigh-Jean portion of the spectrum,
so that nearby thermally dominated pulsars (e.g. J1856$-$3754), may be very bright in the
soft X-ray but have modest ionizing flux. In contrast, the power-law magnetospheric flux provides
a large photon flux near 13.6\,eV.  2PC includes a compilation of $0.3-10$\,keV fluxes
for the LAT pulsars so for convenience and uniformity we estimate ${\dot N}_{\rm ion}$ from these 
data. Although more detailed sums can be made for individual well-studied pulsars, this  provides
convenient assessment of the survey targets. For consistency with 2PC, we assume $kT \sim 50$\,eV
for the thermal emission and $\Gamma =2$ for the non-thermal power law index. With these assumptions
one converts from the unabsorbed $0.3-10$\,keV fluxes, obtaining
$$
\int {\dot N_I} (E) \alpha_{\rm ion} (E){\rm d}E =
\begin{cases}
1.8\times 10^{22} F_{X,-13} d_{kpc}^2 {\rm cm^2s^{-1}}\\
4.9\times 10^{23} F_{X,-13} d_{kpc}^2 {\rm cm^2s^{-1}}
\end{cases}
$$
and $$
\theta_I =
\begin{cases}
0.8^{\prime\prime} F_{X,-13}/\mu_{\rm masy} & {\rm 50\,eV\, Thermal}\\
22^{\prime\prime}  F_{X,-13}/\mu_{\rm masy} & \Gamma=2 \, {\rm PL}
\end{cases}
\eqno (18)
$$
We should exclude the large scale $\gg \theta_a$ downstream PWN flux from the power law
component, but this is not always possible without high resolution (e.g. {\it CXO}) PWN images. 
When $\theta_I> \theta_a$ we may expect pre-ionization and an H$\alpha$ halo. 
PSR J1509$-$5850 and J2030+4415 have the largest $\theta_I/\theta_a$ in Table 1.

If $\theta_I$ is very large, say $\ge 20^{\prime\prime}$, then we would expect an extended halo with
low surface brightness and very little neutral hydrogen reaching the termination shock. This would
compromise H$\alpha$ detection.
Evidently, for J1509$-$5850 this pre-ionization is not complete, since we detect edge brightening 
around the PWN cavity, indicating excitation and/or charge exchange of remnant neutrals in the compressed 
ISM. Indeed the large value for $r_{apex}$ in Table 1 implies small upstream ionization. 
However, with a slightly smaller extinction $A_R$ or smaller pulsar distance, the measurements
are consistent with substantial upstream ionization. Clearly better images are needed to 
obtain accurate bow shock fluxes and remnant HI densities for this novel system.
 
\begin{figure}[ht!!]
\vskip 8.2truecm
\includegraphics{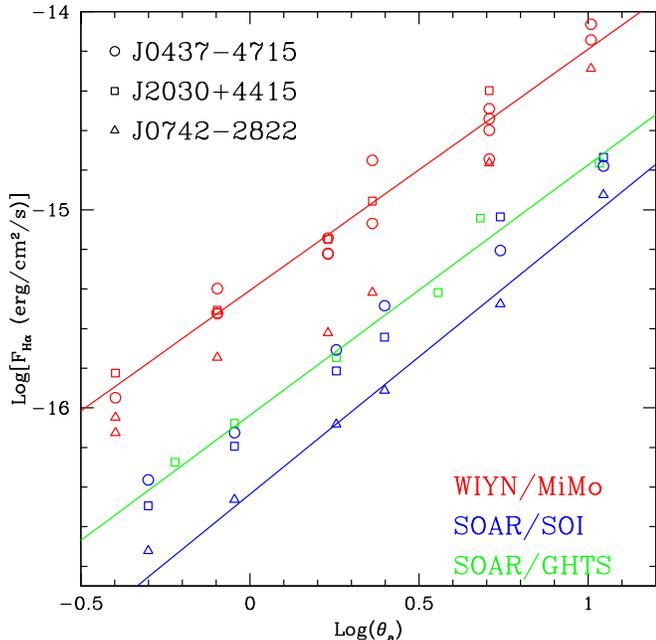}
\begin{center}
\caption{\label{ULcal} Upper limits on undetected bow shocks as a function of standoff scale. 
Three different bow shocks are
used as test templates. The short WIYN/MiMo integrations were read-noise limited. SOAR
observations reached larger depth and were limited by background or electronic noise.
}
\vskip -0.5truecm
\end{center}
\end{figure}

\subsection{H$\alpha$ Upper Limits}

	We have established that we can detect bow shocks appreciably fainter than those 
previously reported. However, to extract conclusions from our survey we must both quantify
the upper limits for the individual targets and compare these limits with the H$\alpha$ flux
expected from these particular sources. This should improve our understanding of the rare 
bow shock phenomenon, helping future campaigns to discover more examples.
With multiple telescope/camera configurations and varying conditions, it is best to
quote limits for individual objects. As expected, our limits are essentially surface brightness 
values. However, for comparison with the sums in \S2 we wish to limit the apex flux. 

	In general, a bow shock nebula stands out {\it via} its bilaterally symmetric
wedge shape. The detectability thus depends somewhat on the nebula geometry. To test this
we selected three nebulae, with apex shapes following Equations (5) and (6), scaled these to a range of
standoff angular size $\theta_a$ and apex fluxes $F_{H\alpha}$, added them to actual survey images
and, via inspection, found the limiting flux below which they would not be noticed. This is
perforce somewhat subjective, but clear trends emerged even though the full bow shock shapes
varied appreciably from PSR J0437$-$4715 (apex only), through PSR J2030+4415
(apex plus a single bubble) to PSR J0742$-$2822 (long low-surface brightness wedge). Figure
6 shows this limiting detection flux as a function of $\theta_a$. As expected, when $\theta_a$ is
small, the surface brightness increases and the bow shock can be discerned to fainter $F_{aH\alpha}$. 
The sensitivity depended slightly but not dramatically on the actual shock shape. We can parametrize
this limit as
$$
F_{aH\alpha} > A \theta_a^{1.3}, \quad \theta_a \ga  0.5^{\prime\prime}, 
\eqno(19)
$$
constant for smaller $\theta_a$. Note that this is shallower than the $\theta_a^2$ for a constant
surface brightness limit. The coherent limb of well-resolved nebulae permits detection
at lower surface brightness.

	The prefactor $A$ depends on the effective noise in the individual image, particularly on the
read-noise, the star field crowding and the presence of diffuse H$\alpha$. The shorter exposure images
(esp. WIYN/MiMo) were read-noise dominated. Longer exposures were limited by crowding or simple S/N.
We quantified this by measuring the limiting flux for $\theta_a=0.8-0.9^{\prime\prime}$
in each image. For consistency we injected the scaled bow shock of PSR J2030+4415; the test was
made at the pulsar position. In most cases this is known to sub-arcsecond precision in our registered images.
In addition, when a proper motion was available, we looked for a nebula aligned with this symmetry axis.
Tests show that these limits are not strongly dependent on  limb shape but are sensitive to the
image noise at the pulsar position. In addition the sensitivity for very large $\theta_a > 10^{\prime\prime}$
nebulae was found to be slightly decreased when the image had structured diffuse emission.

	To convert to an intrinsic flux we need to correct for interstellar extinction. We have
used several resources to estimate this value. As an upper limit, we use the full Galactic extinction
from \citet{sf11}, as computed by the NASA Extragalactic Database (NED). For many of our sources,
there are extinction estimates listed in 2PC, determined by converting an $N_H$ estimate
from X-ray spectroscopy. In a few cases we supplement these by literature-measured $N_H$ and 
follow 2PC in converting to $A_R = N_H/2.2 \times 10^{21} {\rm cm^{-2}}$. For radio
pulsars \citet{hnk13} have computed a $DM -N_H$ correlation, which gives $A_R=0.013 DM/{\rm cm^{-3}pc}$.
For the inner Galaxy, we can use the $A_{K_s}(d)$ compilation of \citet{met06} to estimate 
the extinction to our targets at the adopted distance $d$. In many cases this provides the only
estimate for $\gamma$-ray discovered, non-radio PSR. For consistency
we adopt the minimum of the various estimates, to provide a lower limit
to the foreground extinction, while avoiding bias from intrinsic absorption, etc. The corresponding
$A_R$ values, appropriate for H$\alpha$, are listed in Table 2.

	We did detect stellar continuum sources at the positions of 15 of our binary targets. As it happens,
all have previously been reported in the literature, so we do not detail these detections here.

\begin{deluxetable*}{lrrrrrrrrrrrrr}[h!!]
\tablecaption{\label{HaUL} Pulsars with Flux Upper Limits}
\tablehead{
\colhead{Name}& \colhead{$t_{\rm exp}^a$}& \colhead{$f_{\rm obs-5}^b$}& \colhead{$d$/Type$^c$}& \colhead{$A_R$} &\colhead{$v_\perp$}
& \colhead{${\dot E}_{34}$} &\colhead{$f_{\rm mod-5}^d$} &\colhead{rat$^d$} &\colhead{$\theta_{mod}$} &\colhead{$\theta_I$} 
&\colhead{$n_{\rm WNM}^e$} &\colhead{$\phi_{\rm WNM}^e$} &\colhead{Flg$^f$}           \cr
              & s Tel              & & kpc\quad  & mag              &km/s
&             & &                &$^{\prime\prime}$ &$^{\prime\prime}$
&${\rm cm^{-3}}$   &
}
\startdata
J0023+0923   & 600 G&  4.1&  0.69 D& 0.19& 100&  2.87&   174.9&   0.234&  5.32&    0.15& 0.13& 0.16& 0.6\\
J0030+0451   &1200 G&  2.6&  0.32 P& 0.04&   8.7&  0.65&   8331.&   0.050& 45.29&   10.07& 0.26& 0.32& 0.0\\
J0034$-$0534 &1200 G&  2.1&  0.54 D& 0.08&  79.5&  2.94&   453.8&   0.081&  8.17&    0.01& 0.15& 0.18& 1.0\\
J0101$-$6422 &1200 G&  2.4&  0.55 D& 0.05&  40.8&  1.75&   730.2&   0.085& 11.18& $<$3.25& 0.18& 0.21& 0.0\\
J0102+4839   &1200 W&  8.1&  2.30 D& 0.22& 100&  3.39&   17.99&   1.076&  1.76& $<$0.41& 0.13& 0.15& 0.0\\
J0106+4855   & 600 W&  5.9&  3.00 D& 0.31& 100&  3.82&   10.94&   1.255&  1.73& $<$2.64& 0.09& 0.12& 0.0\\
J0218+4232   & 300 W& 11.8&  2.70 D& 0.12&  64.1& 46.29&   381.0&   0.861& 11.65&   20.33& 0.07& 0.12& 0.0\\
J0248+6021   & 300 W& 58.8&  2.00 K& 3.52& 646.0& 27.61&    0.57&   $>$99&  0.47& $<$2.91& 0.47& 0.35& 0.0\\
J0307+7443   & 600 W& 11.0&  0.60 D& 0.08& 100&  4.07&   361.2&   0.246&  4.48&      --& 0.36& 0.37& 0.6\\
J0340+4130   & 600 W&  6.6&  1.80 D& 0.22& 100&  1.23&   10.69&   0.753&  1.05& $<$0.38& 0.22& 0.27& 0.6\\
J0357+3205   & 300 W&  9.6&  0.60 G& 0.35& 100&  0.77&   52.64&   0.511&  2.00&    2.75& 0.34& 0.37& 0.0\\
J0533+6759   & 300 W& 11.8&  2.40 D& 0.65& 100&  1.09&    3.59&   4.579&  1.16&      --& 0.09& 0.12& 0.0\\
J0554+3107   & 300 W&  6.8&  2.07 G& 1.18& 100&  7.28&   19.80&   0.756&  1.65&      --& 0.40& 0.37& 0.6\\
J0605+3757   & 600 W& 11.0&  0.70 D& 0.27& 100&  1.80&   98.40&   0.403&  2.41&      --& 0.40& 0.37& 0.6\\
J0610$-$2100 & 600 W&  7.3&  3.50 D& 0.15& 302.6&  0.19&    0.09&   $>$99&  0.17& $<$1.39& 0.04& 0.11& 0.0\\
J0613$-$0200 & 600 W&  6.6&  0.90 P& 0.11&  46.0&  2.31&   284.1&   0.212&  4.92&    1.96& 0.35& 0.37& 0.0\\
J0614$-$3329 & 600 S&  4.2&  1.90 D& 0.09& 100&  4.27&   37.55&   0.503&  2.84&    2.80& 0.09& 0.12& 0.6\\
J0622+3749   &1200 W&  4.8&  1.89 G& 0.44& 100&  3.53&   22.72&   0.487&  1.72& $<$5.09& 0.21& 0.26& 0.6\\
J0631+1036   & 300 W& 10.3&  6.50 O& 0.87& 100& 22.51&    8.23&   1.203&  0.87& $<$1.53& 0.44& 0.36& 0.0\\
J0633+0632   & 300 W& 11.0&  1.06 G& 0.34& 100& 15.54&   350.9&   0.239&  4.29&    3.91& 0.48& 0.35& 0.6\\
J0729$-$1448 & 600 W&  6.6&  3.50 D& 1.20& 100& 36.69&   34.17&   0.605&  2.16& $<$1.33& 0.41& 0.37& 0.6\\
J0734$-$1559 & 900 W&  5.1&  1.40 G& 0.88& 100& 17.21&   133.8&   0.227&  3.53& $<$3.46& 0.44& 0.36& 0.6\\
J0737$-$3039 & 360 G& 16.5&  1.10 P& 0.63&  23.0&  1.13&   162.3&   1.004&  5.24&      --& 0.41& 0.37& 0.0\\
J0751+1807   & 600 W&  5.1&  0.40 P& 0.12&  11.4&  1.36&   6858.&   0.085& 34.25&    2.21& 0.35& 0.37& 0.0\\
J0908$-$4913 &1200 S&  3.8&  2.50 D& 2.34& 100& 63.74&   40.53&   0.610&  3.79& $<$1.03& 0.45& 0.36& 0.6\\
J1016$-$5857 &1200 S& 29.6&  2.60 K& 1.19& 100& 334.0&   568.4&   1.005&  8.77&   13.56& 0.41& 0.37& 0.0\\
J1019$-$5749 & 600 S& 12.7&  6.90 D& 6.61& 100& 24.00&    0.04&   $>$99&  0.88& $<$10.8& 0.41& 0.37& 0.0\\
J1023$-$5746 &1200 S& 27.5&  2.26 G& 0.65& 100& 1417.&   5226.&   0.279& 19.09&    4.25& 0.48& 0.35& 0.0\\
J1024$-$0719 &1800 W&  4.4&  0.53 D& 0.08& 150.8& 0.09 &    5.36&   0.515&  0.63&    0.01& 0.22& 0.27& 1.0\\
J1028$-$5819 & 600 S& 29.6&  2.30 D& 0.66& 100& 108.3&   383.6&   0.759&  5.22&    1.08& 0.48& 0.35& 0.6\\
J1044$-$5737 &1200 S&  5.3&  1.32 G& 0.24& 100& 104.4&   1655.&   0.064&  9.01&    1.27& 0.47& 0.35& 0.6\\
J1048$-$5832 & 600 S& 12.7&  2.90 D& 1.68& 100& 259.6&   226.5&   0.730&  6.48&    3.32& 0.47& 0.36& 0.6\\
J1105$-$6107 & 300 S& 38.1&  4.90 D& 2.20& 100& 322.1&   60.88&   5.072&  4.50& $<$0.41& 0.42& 0.37& 0.0\\
J1119$-$6127 & 600 S& 14.0&  8.40 K& 5.11& 100& 302.9&    1.34&   40.70&  2.56&   19.48& 0.41& 0.37& 0.0\\
J1124$-$3653 &1200 S&  3.0&  1.70 D& 0.22& 100&  3.08&   29.99&   0.382&  2.55&    0.80& 0.10& 0.13& 0.6\\
J1125$-$5825 & 600 S& 10.2&  2.60 D& 0.64& 100& 15.34&   43.34&   0.642&  1.95&      --& 0.38& 0.37& 0.6\\
J1135$-$6055 & 300 S& 16.9&  3.01 G& 0.82& 100& 268.3&   478.6&   0.449&  6.36&    7.05& 0.46& 0.36& 0.0\\
J1142+0119   &1200 W&  6.6&  0.90 D& 0.04& 100&  0.86&   35.14&   0.867&  2.91&      --& 0.08& 0.12& 0.6\\
J1231$-$1411 & 600 W&  4.8&  0.40 D& 0.11& 118.2&  1.11&   168.5&   0.171&  3.58&    1.46& 0.25& 0.31& 1.0\\
J1301+0833   & 900 W&  2.8&  0.70 D& 0.06& 100& 12.65&   840.0&   0.101& 12.60&      --& 0.10& 0.13& 0.6\\
J1311$-$3430 & 600 W&  3.4&  1.40 D& 0.14&  53.2&  9.10&   362.3&   0.215& 10.02&    6.60& 0.10& 0.13& 1.0\\
J1312+0051   &1200 W& 10.3&  0.80 D& 0.07& 100&  1.70&   85.66&   0.915&  4.29&      --& 0.09& 0.12& 0.6\\
J1410$-$6132 & 600 S&  9.3& 15.60 D& 8.81& 100& 1300.&    0.06&   $>$99&  2.68& $<$29.3& 0.47& 0.35& 0.0\\
J1413$-$6205 & 600 G& 11.4&  1.32 G& 0.73& 100& 106.4&   1071.&   0.211&  8.97&    1.95& 0.48& 0.35& 0.6\\
J1418$-$6058 &1800 G&  4.0&  1.49 G& 1.22& 100& 642.8&   3224.&   0.066& 19.20&    0.56& 0.50& 0.34& 0.6\\
J1420$-$6048 & 300 S&  5.1&  5.60 D& 4.66& 100& 1342.&   20.08&   4.033&  7.57&    9.36& 0.47& 0.35& 0.0\\
J1422$-$6138 &1200 S&  5.1&  1.64 G& 0.86& 100& 12.54&   72.72&   0.262&  2.49&      --& 0.48& 0.35& 0.6\\
J1429$-$5911 & 600 S&  6.8&  4.08 G& 3.52& 100& 100.7&    8.13&   4.146&  3.09& $<$3.11& 0.40& 0.37& 0.0\\
J1446$-$4701 &1800 S&  2.5&  1.50 D& 0.34& 100&  7.04&   78.64&   0.144&  2.84& $<$2.35& 0.25& 0.31& 0.6\\
J1459$-$6053 & 600 S& 21.2&  1.50 G& 1.30& 100& 118.2&   548.7&   0.732&  8.65&    1.33& 0.45& 0.36& 0.6\\
J1514$-$4946 & 600 S&  5.9&  0.90 D& 0.41& 100&  3.03&   88.44&   0.250&  2.48& $<$0.15& 0.39& 0.37& 0.6\\
J1522$-$5734 & 600 S&  7.6&  2.29 G& 2.34& 100& 150.5&   114.2&   0.812&  6.15&      --& 0.48& 0.35& 0.6\\
J1531$-$5610 & 600 S&  6.8&  2.10 D& 1.44& 100& 118.6&   245.6&   0.314&  5.85&    3.56& 0.50& 0.34& 0.6\\
J1536$-$4989 & 600 G&  8.9&  1.80 D& 0.98& 100&  2.57&   11.07&   1.167&  1.20&      --& 0.35& 0.37& 0.0\\
J1544+4937   & 300 W&  3.7&  1.20 D& 0.04& 100&  2.32&   53.63&   0.521&  4.29&      --& 0.06& 0.11& 0.6\\
J1551$-$0658 & 900 W&  2.4&  1.00 D& 0.28& 100&  0.21&    5.60&   0.502&  1.02&      --& 0.13& 0.15& 0.6\\
J1600$-$3053 &1200 G&  2.7&  2.40 D& 0.43&  82.1&  1.35&    7.34&   0.712&  1.49&    0.02& 0.10& 0.13& 1.0\\
J1620$-$4927 & 600 S& 29.6&  0.74 G& 0.58& 100& 10.60&   388.0&   0.705&  4.97& $<$0.52& 0.49& 0.34& 0.6\\
J1630+3734   &1200 W&  3.4&  0.90 D& 0.02& 100&  2.24&   93.37&   0.245&  3.90&      --& 0.12& 0.14& 0.6\\
J1648$-$4611 & 600 S&  5.1&  4.90 D& 4.41& 100& 27.15&    0.67&   12.22&  1.30& $<$1.11& 0.43& 0.37& 0.0\\
J1658$-$5324 & 900 S&  8.5&  0.90 D& 0.40& 100&  5.68&   166.9&   0.283&  3.38&    1.17& 0.39& 0.37& 0.6\\
J1713+0747   & 900 W&  5.9&  1.10 P& 0.21&  32.9&  0.63&   78.81&   0.574&  4.33& $<$6.29& 0.16& 0.20& 0.0\\
J1718$-$3825 & 600 S& 25.4&  3.60 D& 1.78& 100& 162.5&   83.76&   2.192&  4.12&    9.44& 0.47& 0.35& 0.0\\
J1730$-$3350 & 600 G&  7.9&  3.50 D& 3.37& 100& 160.4&   20.28&   2.804&  4.10& $<$0.21& 0.49& 0.34& 0.0\\
J1741+1351   & 600 W&  3.9&  0.90 P& 0.22&  50.1&  4.14&   405.6&   0.151&  7.55& $<$4.32& 0.23& 0.28& 1.0\\
J1745+1017   & 300 W& 11.0&  1.30 D& 0.31&  48.2&  0.81&   37.10&   1.238&  2.70&      --& 0.18& 0.22& 0.0\\
\enddata
\end{deluxetable*}

\setcounter{table}{1}
\begin{deluxetable*}{lrrrrrrrrrrrrr}[ht!!]
\tablecaption{\label{HaUL} Pulsars with Flux Upper Limits (Continued)}
\tablehead{
\colhead{Name}& \colhead{$t_{\rm exp}^a$}& \colhead{$f_{\rm obs-5}^b$}& \colhead{$d$/Type$^c$}& \colhead{$A_R$} &\colhead{$v_\perp$}
& \colhead{${\dot E}_{34}$} &\colhead{$f_{\rm mod-5}^d$} &\colhead{rat$^d$} &\colhead{$\theta_{mod}$} &\colhead{$\theta_I$} 
&\colhead{$n_{\rm WNM}$} &\colhead{$\phi_{\rm WNM}$} &\colhead{Flg$^e$}           \cr
              & s Tel              & & kpc\quad  & mag              &km/s
&             & &                &$^{\prime\prime}$ &$^{\prime\prime}$
&${\rm cm^{-3}}$   &
}
J1809$-$2332 & 600 W& 12.5&  1.70 K& 1.57& 218.0& 55.88&   48.69&   0.937&  2.44&   12.73& 0.43& 0.37& 0.0\\
J1810+1744   & 600 W&  9.2&  2.00 D& 0.14& 100&  7.59&   57.42&   0.795&  3.09&    0.38& 0.13& 0.15& 0.6\\
J1813$-$1246 & 300 G&  3.7&  1.73 G& 0.35& 100& 811.8&   6752.&   0.031& 20.26&    2.48& 0.42& 0.37& 0.6\\
J1826$-$1256 & 300 W& 22.0&  1.22 G& 0.61& 100& 465.5&   6103.&   0.204& 20.06&    3.36& 0.49& 0.35& 0.6\\
J1833$-$1034 & 300 W&  8.1&  4.30 K& 2.20& 100& 4376.&   1071.&   0.391& 18.76& $>$99.9& 0.43& 0.37& 0.0\\
J1835$-$1106 & 300 W& 23.5&  2.80 D& 1.72& 824.6& 23.13&    0.88&   $>$99&  0.26& $<$0.10& 0.42& 0.37& 0.0\\
J1838$-$0536 & 600 W&  7.0&  1.98 G& 1.40& 100& 771.7&   1863.&   0.157& 15.93&    1.83& 0.49& 0.35& 0.0\\
J1846+0919   & 300 W& 14.0&  1.53 G& 0.88& 100&  4.44&   29.15&   1.215&  1.84& $<$4.66& 0.36& 0.37& 0.0\\
J1902$-$5105 &1200 G&  2.4&  1.20 D& 0.13& 100& 12.82&   271.9&   0.098&  5.79& $<$0.71& 0.17& 0.20& 0.6\\
J1907+0602   & 600 G&  2.3&  1.42 G& 1.07& 100& 367.3&   2360.&   0.039& 15.65&    0.85& 0.47& 0.35& 0.6\\
J1954+2836   & 600 W&  5.5&  1.74 G& 0.66& 100& 136.4&   848.5&   0.106&  7.68& $<$1.37& 0.49& 0.35& 0.6\\
J1957+5033   & 300 W& 10.3&  0.91 G& 0.25& 100&  0.67&   21.80&   0.715&  1.24& $<$0.31& 0.33& 0.37& 0.6\\
J2017+0603   & 300 W&  9.6&  1.60 D& 0.31& 100&  2.55&   25.81&   0.980&  1.90&    0.18& 0.17& 0.21& 0.0\\
J2021+4026   & 600 W& 29.4&  1.50 K& 1.57& 100& 14.89&   53.65&   2.727&  3.09&    0.28& 0.44& 0.36& 0.0\\
J2028+3332   & 600 W&  8.8&  0.99 G& 0.44& 100&  4.52&   106.5&   0.326&  2.58& $<$1.62& 0.44& 0.36& 0.6\\
J2043+1711   & 600 W& 14.7&  1.80 D& 0.16& 111.1&  2.19&   17.16&   1.618&  1.47& $<$1.66& 0.16& 0.19& 0.0\\
J2051$-$0827 & 600 G&  3.0&  1.00 D& 0.19&  34.7&  1.00&   140.9&   0.244&  5.92&    0.01& 0.15& 0.18& 0.0\\
J2111+4606   & 300 W& 13.2&  4.90 D& 1.32& 100& 186.8&   79.41&   1.025&  3.64& $<$11.5& 0.37& 0.37& 0.0\\
J2129$-$0429 &1200 G&  3.5&  0.90 D& 0.09& 100&  4.04&   157.8&   0.191&  4.76&      --& 0.14& 0.16& 0.6\\
J2139+4716   & 300 W& 14.7&  0.86 G& 0.44& 100&  0.41&   12.71&   1.157&  0.90& $<$2.87& 0.43& 0.37& 0.0\\
J2214+3002   & 300 W& 10.3&  1.54 D& 0.16& 100&  3.71&   46.64&   0.952&  2.77&    1.20& 0.13& 0.15& 0.0\\
J2215+5135   & 300 W& 14.7&  3.01 D& 0.18& 100&  9.87&   31.78&   0.918&  1.52&    2.42& 0.30& 0.35& 0.0\\
J2234+0944   & 300 W& 13.2&  0.80 D& 0.24& 100&  3.08&   133.2&   0.824&  4.58&      --& 0.15& 0.17& 0.6\\
J2238+5903   & 300 W& 16.9&  2.11 G& 3.08& 100& 115.6&   52.25&   3.691&  5.86& $<$9.91& 0.48& 0.35& 0.0\\
J2240+5832   & 300 W& 19.1&  7.70 O& 3.08& 801.0& 28.55&    0.04&   $>$99&  0.10& $<$4.62& 0.48& 0.35& 0.0\\
J2241$-$5236 &1200 G&  2.4&  0.50 D& 0.03& 100&  6.22&   831.0&   0.058&  9.14&    0.27& 0.19& 0.23& 0.6\\
J2256$-$1024 & 300 W& 12.5&  0.60 D& 0.09& 100&  9.83&   862.9&   0.366& 10.80&      --& 0.15& 0.18& 0.6\\
J2302+4442   & 300 W&  8.8&  1.20 D& 0.18& 100&  0.71&   14.47&   0.810&  1.12&    0.85& 0.25& 0.31& 0.6\\
J2339$-$0533 &2400 G&  2.5&  0.40 D& 0.07& 100&  4.21&   850.3&   0.056&  8.81&    0.96& 0.21& 0.27& 0.6\\
\enddata
\tablenotetext{}{Tabulated quantities, in order: Pulsar name, W012 exposure and camera, measured H$\alpha$ flux
limit at fixed scale, distance and method, estimated extinction, adopted perpendicular velocity, 
spindown power,
expected flux for the WNM model, ratio of observed limit ito model flux at the model angular scales, model angular scale,
ionization angular scale, model local WNM density, model local WNM fill factor, 
detectability flag. See text for details. 
}
\tablenotetext{a}{Letter denotes Telescope/camera: G=SOAR/GHTS, S=SOAR/SOI, W=WIYN/MiMo}
\tablenotetext{b}{H$\alpha$ flux upper limit for J2030+4415 template at $\theta_a= 0.9^{\prime\prime}$ scale 
in units of $10^{-5} {\rm cm^{-2}s^{-1}}$.}
\tablenotetext{c}{Distance Type codes are from 2PC; D=DM, G=$\gamma$-ray (Equation 16), K=kinematic, O=other, P=parallax}
\tablenotetext{d}{Model flux in  $10^{-5} {\rm cm^{-2}s^{-1}}$; rat=$f_{\rm obs-5}$(scaled to $\theta_{mod}$)/$f_{\rm mod-5}$.}
\tablenotetext{e}{$n_{\rm WNM}$ and $\phi_{\rm WNM}$ are the model WNM density and fill factor at the pulsar $Z$.} 
\tablenotetext{f}{Flg=0 when known parameters imply no H$\alpha$ bow shock; 1.0 when when $v_\perp>100$km/s, 0.6 otherwise.}
\end{deluxetable*}

\begin{figure}[ht!!]
\vskip 8.1truecm
\includegraphics{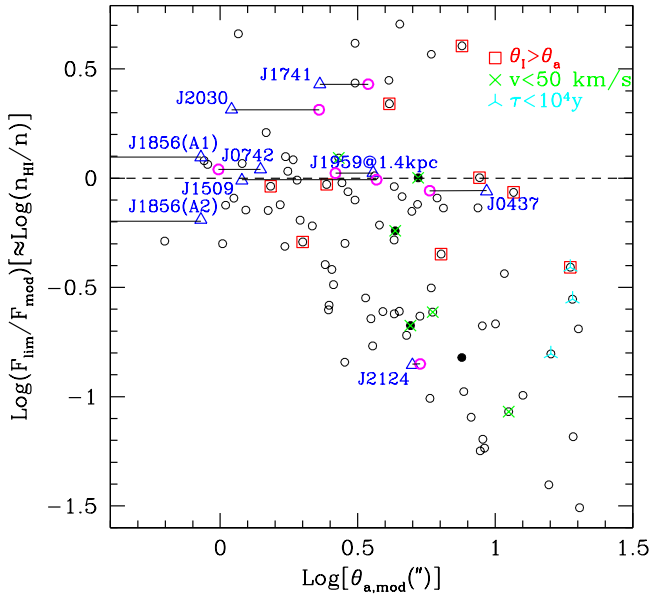}
\begin{center}
\caption{\label{ULs} 
H$\alpha$ survey flux upper limits (as a fraction of the apex model flux, assuming residence in the WNM) 
plotted against the model apex scale $\theta_a$. Points lying below the dashed line give 
meaningful upper limits on the local neutral fraction; low values suggest the pulsar
lies in an ionized medium (eg. WIM).  Solid points have parallax distances. Overlays on survey points 
indicate various conditions counter-indicating a bright bow shock (squares: likely upstream
photoionization, x: velocity likely too low for a strong shock, 3-point star: a young
pulsar likely close to ionizing SNR or parent association).
For comparison, triangles show the detected bow shocks at the observed $\theta_a$, lines connect to the
model $\theta_a$ estimates at typical WNM density; the vertical position indicates the apex flux 
relative to the model at the Table 1 $d$.
Note that most observations give significant sensitivity at the modeled flux. However
PSRs J0742$-$2822 and J2124$-$3358 lie relatively close to the survey sensitivity boundary; 
scaled with distance, we would expect to detect nebulae like these $\sim 30\%$ of the time.
We have little sensitivity to pulsars like J1856$-$3754 (plotted for both Case A1 and Case A2). 
Given the WNM filling factor
and other constraints, we expected to detect $\sim 9$ sources. Seven LAT pulsars are detected;
our survey has contributed three.
}
\vskip -0.5truecm
\end{center}
\end{figure}

\subsection{Detection Statistics and the HI in the ISM}

	To detect an $ H\alpha$ bow shock the pulsar must reside in partly neutral 
ISM. Thus to compare our measured upper limits with an expected bow shock flux, we must
estimate the density and filling fraction of the neutral phases of the ISM. For the
cold neutral medium (CNM) we assume high density, resulting in a small volume filling
factor. The most important phase for our comparison is the warm neutral medium (WNM)
which has a mid-plane density $\sim 0.3-0.6 {\rm cm^{-3}}$ and volume filling factor
$\phi \sim 0.35$. We assume that this is in approximate pressure 
equilibrium with the warm ionized medium (WIM) and, accordingly, twice
the H space density. \citet{kh87} have used dispersion measure and emission measure
observations of the WIM to derive the density and filling factor run with Galactic
height $Z$. Following their arguments, we can note that WNM and WIM temperatures are 
very similar so that using their expression for $\phi_{WIM}(Z)$, but updating the
HI structure to that of \citet{fer01} we obtain
$$
{\rm CNM}: n_{\rm CNM} = 30,\qquad\qquad  \phi_{\rm CNM} =0.013 e^{-(Z/0.13)^2}
$$
$$
{\rm WNM}: n_{\rm WNM} = 0.5\,e^{-|Z|/0.43},\qquad\qquad\qquad\qquad\eqno(20)
$$
$$
\phi_{\rm WNM} =
	0.57\,(0.19\,e^{-(Z/0.32)^2} + 0.11\,e^{-|Z|/0.4})/n_{WNM}
$$
$$
{\rm WIM}: n_{\rm WIM} = 0.25\,e^{-|Z|/0.43},\qquad  \phi_{\rm WIM} = 0.1\,e^{|Z|/0.75}
$$
with scales in kpc and number densities in ${\rm cm^{-3}}$. We assume ionization fractions
of 0.0, 0.05 and 0.95 for these three phases \citep{set00}. \citet{kh87} discuss the possibility that the
WIM is over-pressured with respect to the WNM, which would then halve the WNM density and double
its fill factor.

	We now apply the modeling of \S2 to our survey observations. For each pulsar
we use the distance and plane height $Z=d\, {\rm sin}( b)$ to compute the local WNM density and
filling factor (Eqs. 20), the resulting apex standoff angle $\theta_a$ (Eq. 4) and the expected apex flux 
$f_{\rm H\alpha}$ (Eq. 10), using measured pulsar parameters whenever possible.
The model H$\alpha$ flux is subject to our estimated extinction \S4.5. This number is
then compared with our measured upper limit, scaled (Equation 19) for our estimated $\theta_a$.
We tabulate the resulting ratio in Table 2.  When this ratio is $<1$, we have adequate
sensitivity to detect the predicted bow shock flux. 

	We can also use the cataloged X-ray fluxes (from 2PC) to estimate the ionization
angle (Eq. 18). We compute for both thermal and non-thermal fluxes; the non-thermal
emission nearly always dominates and the combined $\theta_I$ is listed in Table 2. 
Even when we have enough sensitivity to detect a bow shock, an H$\alpha$ nebula may
not exist. In particular, we require a strong shock. Here we take $v_\perp > 50{\rm km/s}$,
recognizing that some pulsars may have ${\rm sin}i \ll 1$ and shock with small transverse velocity.
Of the $\sim 100$ objects covered in this survey only 20 had measured proper motions.
Although pulsars are a high velocity population with $v_{2D} \approx 300{\rm km\,s^{-1}}$,
we were surprised to find that 8 of 20 had $v_\perp <  50{\rm km/s}$. Thus we conservatively assume
that as many as 40\% of the pulsars without parallax measurements may be too slow to produce
robust H$\alpha$ bow shocks. In addition, very young pulsars lie within their parent supernova
remnants and have not escaped to the external neutral medium. We assume that pulsars
with ${\rm log \tau}<4$ are unlikely to show bow shocks. Finally if $\theta_I>\theta_a$
the ISM may be sufficiently pre-ionized to render a bow shock undetectable. Combining
these factors, if a pulsar has ${\rm log \tau}>4$, $\theta_a>\theta_I$ and $r<1$, 
$P_{det}=\phi_{WNM}$ if a measured proper motion gives $v_\perp > 50{\rm km/s}$,
$P_{det}=0.6 \phi_{WNM}$ if $\mu_T$ is unmeasured. $P_{det}=0$ otherwise. For
our survey $\Sigma P_{det}= 9.1$ for the WNM. We expect less than one detected bow shock in
the CNM and WIM. The total number of observed LAT pulsars that satisfy our survey criteria
is seven. 

While we only report three new detections, the other four objects would certainly
have been (prime) targets in our survey, and easily discovered. The 31\% Poisson probability
of seven or fewer detections is not unacceptably small. Systematic effects could bias our predicted model flux 
or our survey sensitivity somewhat high. The total numbers are not very sensitive to our 
mass-moment of inertia assumptions; assuming that all pulsars have $M=1.4 M_\odot$ reduces
$\Sigma P_{det}$ to 8.6.  If we under-estimated the extinction or overestimated
our sensitivity by $4\times$ we would bring the number of expected detections to $\sim 3$.
Similarly, increasing all non-parallax distance estimates by $2\times$ would bring the
detections under 4. Alternatively we could assume that pulsars with ${\rm Log(\tau)}< 5.2$
(the youngest bow shock seen) are still in high mass star forming regions with little HI;
this brings the expected number down to 5.5.
However such radical changes seem unlikely, especially since the model 
predicts the observed fluxes well and {\it under-}predictions fluxes for bow shocks with
DM over-estimates.

	One interpretation is that our survey argues against a very large filling fraction for 
the WNM. Our present model implies $\phi_{\rm WNM} =0.37$ at mid-plane, but other pictures
suggest an even larger filling factor. As noted, if the WNM is under-pressured by $2\times$ 
this doubles its filling fraction. The expected bow shock number also doubles, giving a Poisson
probability of seven or fewer detections of 0.25\%, which seems
unacceptable.  A larger H$\alpha$ bow shock sample would help refine these conclusions. With very
high sensitivity, we should also discover bow shocks in the WIM. As Figure 7 shows, at present we
only have a few pulsars observed with sufficient sensitivity for detection in a 95\% ionized WIM.
It will be interesting if deeper surveys, e.g. with 10m-class telescopes, can recover more
of this faint population. The sums above provide guidance to the most favorable targets.

\section{Principal Results and Conclusions}

	In a quest to further understand the rare pulsar H$\alpha$ bow shocks we have observed
a large sample of energetic, {\it Fermi} LAT-detected pulsars; sensitive re-measurement of known examples
provided context for survey imaging of nearly 100 new targets. Three new H$\alpha$ shocks
were discovered, including the detection of two examples of previously unobserved ionization
precursors (PSRs J1509$-$5850 and J2030+4415). 

	To interpret these data we have developed a simple analytic model that 
describes well the apex zone of all known pulsar bow shocks, including conditions likely
to produce precursor halos. This analysis provides a good understanding of the apex zone's
angular scale and flux, treating both the high powered pulsars selected by the LAT and
the low ${\dot E}$ nearby pulsars such as PSR J1856$-$3754 and J2225+6535 which apparently
feature incomplete ionization in the post-shock flow. This exercise provides a quantitative
connection between the bow shock flux and pulsar parameters, applicable over a range 
of, e.g. $10^6\times$ in spindown luminosity. We argue (equation 10) that the flux is independent of ISM density
and highly sensitive to the pulsar distance. When the distance is poorly known, bow shock
observations provide independent measurements of this crucial parameter. We derive
a 0.72\,kpc distance for the $\gamma$-ray only pulsar J2030+4415. We show that the
dispersion measure distances for the millisecond pulsar J1959+2048 and the young, fast
PSR J2225+6535 are too large, deriving $d\approx 1.4$\,kpc and $d=0.8-1$\,kpc for these
two objects. These distance estimates do depend weakly on velocity inclination $i$,
the upstream neutral fraction $f_{HI}$ and, possibly, pulsar wind anisotropy, but these factors can 
be constrained by measurements of the bowshocks' apex shape and emission line profile. Foreground extinction 
also affects the estimate, but this can be determined by measuring the nebula Balmer decrement
or other absorption studies. Thus pulsar bow shocks, while rare, can provide important new
insight into pulsar distances and energetics.

	One aspect of this study deserves special attention. The bow show apex covers a large solid
angle as viewed from the pulsar and thus provides comprehensive measurement of its mechanical spin-down
energy deposition. When a pulsar has a well-measured distance and proper motion, then the bow shock
flux puts a lower limit on ${\dot E}$, which implies a lower limit on the neutron star moment of inertia.
PSR J0437$-$4715 is the premier example. Our apex flux of $(6.7\pm0.7)\times 10^{-3} {\rm H\alpha\,cm^{-2}\,s^{-1}}$
translates to a lower limit of $I_{45} > 1.5$ if the pulsar motion is in the plane of the sky
and upstream medium is 100\% neutral.  Since ${\rm sin}\,i$ and $f_{HI}$ are $\le 1$,
accounting for these factors will only increase this lower bound. There is in addition a small increase
since the photon ($\gamma$-ray) energy losses do not contribute to the mechanical losses in the wind,
but for PSR J0437$-$4715 this accounts for $\sim 1\%$ of ${\dot E}$ (lower luminosity MSP can be 
substantially more $\gamma$-efficient). The extinction is small and adds little
uncertainty to the luminosity of this well-observed bow shock. We have assumed a spherical
pulsar wind, but this can be tested by detailed modeling and measurement of the apex H$\alpha$ velocity structure.
There is one significant assumption in our computation, that the $e^-$ and ions equilibrate (Equation 9);
if we use the often-assumed $\epsilon_{\rm H\alpha}=0.2$, our lower bound on the moment of inertia
would increase to a (likely unphysical) $I_{45} > 3.9$. If the electrons are completely out of equilibrium
the implied moment of inertia is even larger.

Since PSR J0437$-$4715 also has a mass measurement, it should help fulfill the promise of \citet{ls05}
that moment of inertia measurements can pin down the neutron star equation of state. Already
our estimated $I_{45}=1.7\pm 0.2$ prefers relatively stiff EOS. Other pulsar bow shocks, with their near-bolometric
measure of the spin-down power can contribute to this quest. It would be very important to obtain a precise
independent parallax of PSR J1959+2058, since the high pulsar mass $M_{NS} =2.4M_\odot$ implies
a very high moment of inertia $I_{45} = 2.97$, according to Equation (1). If $I_{45}$ is lower, one would require
an even smaller distance than the $d=1.4$\,kpc determined in this paper. Of course, one or more young
neutron stars with low masses and moments of inertia could also be valuable anchors to the equation
of state study. Thus precision parallax measurements of the neutron stars showing H$\alpha$ bow shocks
are strongly motivated ingredients in this fundamental physics quest.

	Our new images also reveal more details about the post-apex, large scale H$\alpha$ emission.
It is now clear that the nebula limb very generally describes a series of undulations
or cavities $10-100\theta_a$ downstream. While in some cases, these appear nearly closed,
other shocks e.g. PSR J0742$-$2822 display oscillation about a more or less conical structure.
Thus the guitar nebula is only the most spectacular example of down-stream bubbles. The details
doubtless depend on accidents of the ISM, but it seems likely that this behavior is a symptom
of instabilities in the post-shock PWN flow \citep{vKI08}. Surprisingly, we find that
with the exception of J1856$-$3754, the total flux of the observed bow shock nebula 
are very similar, despite their being 
detected over a wide range of ${\dot E}$, $v$ and $d$. This calls into question previous
efforts to understand the total nebula flux, and suggests that, unlike the apex flux, this
does not simply track the pulsar properties.

	Finally, our modeling has allowed us to compare our survey flux limits with the expected
sizes and fluxes of potential bow shocks around other pulsars.  This study implies that the 
WNM hosting such bow shocks should occupy no more than $\sim 30$\% of the nearby Galactic plane.
It appears that additional bow shocks await detection, but that deeper exposures are
needed. Our analysis points to a set of high ${\dot E}$, low extinction objects as the favored
targets for future searches, and suggests that intensive study of these nebulae can provide new
insights into pulsar physics.

\medskip

	We thank Matt Stadnik for assistance with the image reductions and the referee 
for a careful reading.
This work was supported in part by NASA grants NIX08AW30G and NIX10AD11G and NNX12AO68G,
along with Chandra Grant GO0-11097X.

\end{document}